\documentclass[12pt]{article}

\usepackage[margin=1in]{geometry}
\usepackage{setspace}
\usepackage{amsmath}
\usepackage{graphicx}
\usepackage{booktabs}
\usepackage{natbib}
\usepackage{hyperref}
\usepackage{subcaption}
\usepackage{tikz}
\usepackage{authblk}
\usepackage{hyperref}
\usepackage{tabularx}

\graphicspath{{figures/}{figures/aggregation/}{figures/gap_pubsec/}}

\title{How Low Can You Go? \\ Validating Administrative Union Records for Substate Estimates of US Union Membership}

\author[1]{John S. Ahlquist\thanks{\url{jahlquist@ucsd.edu}. Professor, School of Global Policy and Strategy. Harrison Burkholder provided research assistance on this project. We thank Henry Kalinowski from OLMS for his assistance with data.  Michael Becher, Mitch Downey, Forest Gregg, Jake Grumbach, and Daniel Stegmuller each provided very helpful comments.}}
\author[2]{Eric Thai\thanks{\url{ericldthai@gmail.com }.}}
\affil[1]{UC San Diego}
\affil[2]{Independent researcher}

\date{4 October 2026}

\begin{document}

\maketitle

\begin{abstract}
Unions are important economic and political actors at the local level, but the primary source on US union membership, the Current Population Survey (CPS), is unreliable below the state level. Researchers have therefore turned to union-reported membership in federal "LM" filings, geolocating unions and aggregating their reported membership to estimate union density for counties, commuting zones, and congressional districts. We ask whether this practice is sound. We develop and defend multilevel regression with poststratification (MrP) on the CPS as a benchmark, then validate LM-based estimates against it from the state level down. LM, CPS, and MrP are highly correlated at the state level, but this correlation is roughly halved at the congressional-district, commuting-zone, and county levels. The breakdown is driven in part by reporting "lumpiness," which produces improbable values, including union densities exceeding 100\%. Missing public sector union members explains only a small part of this divergence. These issues are substantively consequential: in a standard commuting-zone "China shock" regression, swapping density measures changes the import-exposure coefficient enough to alter its magnitude, significance, and sign. We conclude that LM records remain a rich resource for studying unions as organizations, but aggregating LM-reported membership is not a reliable way to measure union density at fine geographic resolution.
\end{abstract}

\newpage
\onehalfspacing
American labor unions are objects of enduring interest across the social sciences.\footnote{Recent reviews from economics, political science, and sociology include \citet{kaplanGovernmentMarketPolitical2025}, \citet{ahlquistLaborUnionsPolitical2017}, and \citet{rosenfeldUSLaborStudies2019}.} Even as unionization has declined, the structure, resources, and geographic distribution of labor organizations remain areas of active research and vital concern to workers, union leaders, policymakers, politicians, and businesses.

In the United States, the Current Population Survey (CPS) is the primary source of data on union membership.  As a large, high-quality survey administered by the US Census Bureau, the CPS has provided the headline government estimates of union density since 1973.  But CPS-derived unionization estimates have several notable problems: they are survey-based and therefore subject to recall and misreporting bias; reported union membership is not linked to a specific union; the CPS misses retired and unemployed union members; and, most relevant for this paper, small sample sizes make CPS estimates unreliable below the state level.

Due to these shortcomings, enterprising scholars have turned to administrative data.  All US unions representing any private sector or federal workers must file annual ``LM forms'' with the Department of Labor's Office of Labor-Management Standards (OLMS).  These administrative reports, available systematically from 2000 onward, represent a rich trove of data about union organizations, including organizational birth and death, location, leadership, finances, and membership \citep{bredaProductivityDynamicsUnion2026, wilmersLaborUnionsActivist2017, downeyCongressionalElectionsUnion2023, holmesDynamicsUnionOrganization2007, mastersFinancialStatusNational2022}.  By filtering to local-level unions, geolocating union offices, and assigning reported members to the same geography as the union office, social scientists have used LM forms to generate estimates of union membership at the Congressional district (CD) \citep{balcazarGlobalizationUnionsRobots2023, becherLocalUnionOrganization2018b, becherReducingUnequalRepresentation2021} commuting zone (CZ)  \citep{downeyCongressionalElectionsUnion2023, becherMachinesWorkersHeterogeneous}, and county \citep{zulloUnionMembershipPolitical2008a} levels.  \citet[appendix A.1]{becherLocalUnionOrganization2018b} show that their LM-based estimates of state-level unionization correspond closely to those derived from the CPS. \citet{bredaProductivityDynamicsUnion2026} show that national union membership reported on LM forms follows a similar trend to that from the CPS. 

LM data come with their own challenges and weaknesses.  Some are well-known: unions representing only state and local public sector workers are exempt from filing LM forms, implying that LM-based estimates should systematically under count union members.\footnote{Whether this is a problem will depend on the nature of the research question.  Existing studies in political science focus on trade and technology shocks that are arguably most relevant for workers in the private sector.}  Others are less remarked upon, such as uncertainty in identifying local union organizations, significant data entry irregularities, and geographic ``lumpiness'' due to large outlier organizations and some unions sharing addresses.  It is not clear whether lumpiness and errors average out in smaller geographies. The consequences may in fact become more severe, because at smaller scales workers are less likely to live and vote in the same geography as their union's office. 

In this paper, we calibrate LM-based estimates of union density at the substate level.  We are able to reproduce strong state-level correspondence between LM and CPS, but, using multilevel modeling with post-stratification (MrP) \citep{gelmanPoststratificationManyCategories1997, ghitzaDeepInteractionsMRP2013, parkBayesianMultilevelEstimation2004} to produce plausible benchmarks, we show that this correspondence breaks down considerably at the county, CD, and CZ levels.  This is, in part, because LM-based estimates are extremely lumpy in ways that do not readily average out, resulting in a variety of pathologies, including impossible estimated union densities above 100\% in some geographies.  We further show that differential public sector employment explains only a modest proportion of these discrepancies.  We demonstrate that this is consequential for applied research by estimating standard, reduced form ``China shock'' regressions at the CZ level \citep{autorChinaShockLearning2016}.  The estimated relationships between import exposure and union density vary substantially--including sign switching and standard errors that would produce variable inference--depending on which estimate of local unionization is chosen and how questionable entries in the LM data are addressed.  We conclude that LM reports remain a valuable resource for a variety of social science and policy questions focusing on union \emph{organizations}, but aggregating LM-reported membership is \emph{not} a reliable method for estimating union density at fine geographic resolution.

This paper makes three major contributions.  First, we highlight underappreciated problems with existing practice.  By focusing on members, we show that aggregating LM-reported data to low-level geographies may be more problematic than previously recognized, providing scope conditions on the appropriate use of LM forms in scholarly and policy research.  We hope to provoke a reconsideration of some claims in the literature, reorient researchers toward other tools when needed, and stimulate additional work that may increase the reliability of LM-derived data in some circumstances (e.g., by linking with emerging databases of collective bargaining agreements or by more precisely restricting the scope of union organizations analyzed).  Second, we present a cross-scale validation architecture that combines government surveys, model-based estimation, and administrative data to shed light on each.  Although this approach is not completely novel, we present the most developed application to date in political economy.  Finally,  the replication archive for this paper provides a cleaned, updated panel of geolocated local union organizations from 2000--2024, including multiple indicators of data quality and corrections for future scholars' reference.     

\section{CPS weaknesses, LM data, and the appeal of localization}
\label{sec:lmdata}

The Current Population Survey has provided the longest consistent time series asking about union membership in the United States.  Since 1983, the union membership question has been asked as part of the Outgoing Rotation Group (ORG) seen by all employed respondents. The Department of Labor relies on the CPS for its regular reports about the labor market, including unemployment and unionization rates across sectors, states, gender, race, and the nation as a whole.  For years, \citet{hirschUnionMembershipCoverage2003} have provided a major service to the scholarly and policy communities by taking the raw CPS data and producing standardized annual estimates of union membership (density) and coverage rates, publicized on Unionstats.com.  In the remainder of this paper, when we refer to state or national-level ``CPS density'' or ``CPS estimates'' we are referring to those reported on Unionstats.   

For all their virtues, the CPS estimates of unionization are not necessarily the ``gold standard''.   \citet[451]{macphersonFiveDecadesCPS2023} summarize the issues with the CPS succinctly, saying 
\begin{quote}
``CPS does not provide fine geographic detail so the analysis of commuting zone or county-level union density is not possible. In addition, there is relatively little information about firms in the CPS, so the natural unit of unionization under the NLRA is not captured by the CPS. Importantly, the CPS only asks the union question of employed workers, so there is no possibility for unemployed workers to stay members of unions (which happens in practice). The absence of administrative data on unions remains a serious obstacle in the US and it would be very helpful if such data were developed and made available.''  \end{quote}
Additional concerns include the truncated CPS time series \citep{farberUnionsInequalityTwentieth2021, romeroBriefExaminationUnion2023} and the fact that CPS, as a survey, suffers from standard concerns about respondent accuracy and increasing non-response \citep{kruegerEvolutionRotationGroup2017}.  

Anticipating Hirsch \& Macpherson's concerns,  scholars found administrative data.  Under the terms of the Labor Management Reporting and Disclosure Act (LMRDA) and the Civil Service Reform Act, all unions representing any private sector, federal, or postal workers must file annual ``LM forms'' with OLMS.  On these forms, unions report information about their location, officers, and finances, with the level of detail determined by the financial size of the union.  Since 2000, LM forms have asked for membership numbers. On LM forms, unions report whoever they consider to be a member, which can include retirees and unemployed members that the CPS would miss. Such data, combined with modern geolocation and mapping tools, can generate a fine-grained description of union organizations and their resources, including members.

\begin{table}[htbp]
\centering
\caption{The CPS and LM forms as sources for US union membership.}
\label{tab:cps_lm_comparison}
\renewcommand{\arraystretch}{1.3}
\begin{tabularx}{\linewidth}{@{}l X X@{}}
\toprule
 & \textbf{CPS} & \textbf{LM forms (OLMS)} \\
\midrule
Type of data
  & Household survey
  & Mandatory administrative filings \\

Collection method
  & Interviewer-administered; union question in ORG
  & Self-reported annual forms filed by unions \\

Unit of observation
  & Individual worker (\& household)
  & Union organization \\

Membership concept
  & Self-identified member on a survey
  & Whoever the union counts as a member \\

Coverage universe
  & All 16+ civilian employed workers
  & Unions for private-sector, federal, postal workers; state/local public-sector--only unions are exempt \\

Linkage to specific union
  & None
  & Union-reported affiliation \\

Geography of reporting
  & Public-use data includes CBSA; estimates reliable at state level
  & Geocoded based on union address to the county, CZ, or CD \\

Time series
  & ORG union series since 1983 
  & Systematic membership reporting since 2000 \\

Sources of bias or error
  & Respondent recall \& misreporting; small samples below the state level
  & Reporting error, organizational ``lumpiness''; ambiguity in identifying locals \\

Known gaps
  & Misses retired and unemployed members; no firm/establishment detail
  & Misses public-sector--only unions; small filers report limited detail \\

Verification of reported figures
  & Census Bureau resources, survey methodology, and weighting
  & LMRDA recordkeeping rules \& false reporting penalties;  OLMS compliance audits (largely directed at financial reporting) \\
\bottomrule
\end{tabularx}
\end{table}

Scholars across the social sciences have exploited LM reports in ways that have opened new empirical terrain.  In political science, Becher, Stegmueller, and co-authors have made the most extensive and innovative use of LM data, generating localized density estimates at the CD and CZ levels to examine the effects of unionization on a variety of legislative and policy outcomes \citep{becherLocalUnionOrganization2018b, becherMachinesWorkersHeterogeneous, becherGlobalCompetitionLocal, becherReducingUnequalRepresentation2021}. But they were not the first to leverage LM forms in research.  In economics, \citet{holmesDynamicsUnionOrganization2007} use LM forms to track the formation and growth of local union organizations. \citet{downeyCongressionalElectionsUnion2023} links local unions to commuting zones and Congressional campaign donations.  \citet{bredaProductivityDynamicsUnion2026} link union locals to CZs to study how entry and exit of locals affects membership and dues revenues.  Industrial relations scholars have also turned to LM data; \citet{wilmersLaborUnionsActivist2017} uses LM forms to study local union expenditures and member dues revenues. \citet{mastersFinancialStatusNational2022} likewise look at union finances over time.  \citet{zulloUnionMembershipPolitical2008a} uses LM forms for 2000 to estimate county-level union membership per voting age adult. \citet{zulloLaborUnionsCharity2011} uses LM forms to calculate union density and the number of union locals at the county level in Michigan.

Rich as the LM data are, many of the works just cited also comment on their challenges and potential weaknesses.  Most frequently mentioned is the fact that unions exclusively representing state and local government employees are exempt from filing LM forms so locals of purely public-sector workers are absent.  More troubling, it is difficult to bound this error as many unions representing public sector workers also have some members in the private sector and so file LM forms that report both groups together.  \citet{becherLocalUnionOrganization2018b} argue that this bias is relatively small. But \citet{reuningPoliticalNatureUnion2026} uses LM forms to generate a sampling frame for a survey of union leaders in Ohio. Concerned about missing unions of public sector workers, he examined collective bargaining agreements in the  State Employee Relations Board.  After deduplication, Reuning's LM-identified locals accounted for only 40\% of the total count of union locals, suggesting that LM data may miss a substantial fraction of local organizations.   Whether this is a concern will depend on the research question at hand.   

Other concerns have also surfaced.  \citet{zulloUnionMembershipPolitical2008a} explicitly mentions problems with large locals, concerns about geolocating members, and spillovers across geographies but nevertheless proceeds to assign members to the county of their union local for the purposes of voting, which is a function of residence. \citet[online appendix]{bredaProductivityDynamicsUnion2026} mention data entry errors and related anomalies in the LM data as well as their process for making corrections.  \citet{downeyCongressionalElectionsUnion2023} discusses the various terms used for ``local union'' and the challenges in making judgements about which entities to count. \cite{becherReducingUnequalRepresentation2021} notice problems with particularly large union locals in New York state, something we confirm below.

Table~\ref{tab:cps_lm_comparison} collects a summary of the differences between CPS and LM data.  There are existing concerns that LM data may (i) miss some union locals and unionized workers; (ii) contain data entry errors and other irregularities.  When focusing on local union organizations and geolocating unions, there are additional concerns: (iii) what to count as a ``local'' union and (iv) the degree of lumpiness in reporting or member spillovers across geographies.   We take these concerns as suggesting that better validation of LM data, especially membership is warranted when researchers attempt to link these members to relatively fine-grained spatial geographies.  Spillovers, errors, and coding decisions that may cancel out when aggregated to the state level may persist at the county or Congressional district level.  

\subsection{Our LM data}
\label{sec:lm-data}

\subsubsection{Identifying local unions}
\label{subsec:lm-identifying}

We downloaded the annual bulk data files for 2000--2024 and, from each yearly release, retained the base form that every reporting union must file, which records the organization's headquarters address, affiliation, and top-line membership count regardless of the union's size or reporting tier (LM-2, LM-3, or LM-4). Amended refilings are resolved to a single record per union-year by keeping the highest amendment number, and unions are linked across years deterministically on the OLMS file number (``LM code''), which is stable over time and requires no probabilistic matching.

Unions in the United States are typically organized in hierarchical, federal organizations.  So-called ``local'' unions are the lowest level, where most members encounter their union.  Nevertheless, the LMRDA requires organizations at all levels to report members, presenting the danger of double- and triple-counting members across levels.  LM filers self-designate their organization type, but these designations are inconsistent and use a variety of similar terms (``local union'', ``local executive council'', ``lodge'', ``chapter'', etc.). In distinguishing local unions from other entities we rely on separate OLMS-provided coding of union organizations.  OLMS assigns each LM filer a ``TLO code'' taking one of three values: international union (TLO~1), intermediate bodies such as regional councils (TLO~2), and local unions (TLO~3). We restrict the panel to TLO~3.\footnote{We confirmed LM filers self-designating as a ``LU'' (local union) are a subset of TLO~3 union locals.   Note that \citet{bredaProductivityDynamicsUnion2026} filter on an ``LU'' self-designation, yielding 21,714 unique union locals as of 2022.  We find 31,958 over the same timeframe.  We thank Henry Kalinowski from OLMS for his assistance in providing updated and comprehensive spreadsheets of all LM filers and their TLO codes through 2024. Michael Becher and Daniel Stegmueller used an earlier version of this spreadsheet and made us aware of these data.} 

\subsection{Geocoding and data errors}
\label{subsec:lm-geocoding}

To place each local in a county, congressional district, commuting zone, and state, we geocode the mailing address reported on the LM form. Addresses were submitted to Esri's ArcGIS Online geocoder using the North America locator, which matched 99.8\% of records; the small remainder were sent to the Google Maps API with center-of-country
and other fallback returns explicitly rejected so that unmatched addresses fail cleanly rather than resolve to a spurious centroid. Of 593{,}498 records, 151 remain ungeocoded. Each resulting coordinate is assigned to geographies by point-in-polygon join against year-specific boundary files, with congressional-district vintages aligned to the
Congress in session in each calendar year.

Similarly, as others have noted, LM-reported membership figures contain data entry and other errors.  We flag for inspection all union membership values that fluctuate by two orders of magnitude or $\pm \infty$ from year to year.  Some of these are easily recognized as data entry or transcription errors (e.g., concatenating values from two adjacent fields in the LM form) that can be corrected with high confidence.  Many result from a missing entry; for these we linearly interpolate membership between the two closest positive values. 

In the end, we produced an unbalanced panel reflecting the fact that some union locals disappear while new ones emerge.  We identify 32,238 unique local unions and 511,994 local-years with reported members from 2000--2024.  Following existing practice, we aggregate the reported members to the geography in which the mailing address of the local is located for the reporting year.

\subsection{Two versions of the panel: \emph{inclusive} and \emph{exclusive}}
\label{subsec:lm-include-exclude}

A local's members are only as geographically meaningful as the assumption that they live, work, and vote near the address in the union's LM form. For large geographies this seems reasonable.  But at finer resolution this becomes a more problematic assumption.  We find that a nontrivial set of filers violate this assumption in ways that might badly distort any sub-state estimates. We therefore compare two versions of the LM data that differ in how we address two decisions: what to do with extremely large and multi-state locals and how to handle multiple locals that share the same mailing address (often a PO box). The \emph{inclusive} version of the LM data applies no filtering on either of these dimensions. The \emph{exclusive} version drops all locals in the top one percent of the membership distribution (pooled across all years), and those that share a headquarters address with at least one other local in the same year. The rationale is best seen through the cases that motivated each choice.

The top-one-percent rule targets filers that OLMS labels as local (TLO~3) but that clearly function as large statewide or multistate unions or even federations. New York State United Teachers (LM code 070-581) is categorized as a local union, but it reports roughly 698{,}000 members. Inspection of its website shows that it is a state federation whose constituent locals file their own LM forms; the New York City-UFT local (LM code 063-924) claims affiliation with the NYSUT and reports 200{,}000 members already appearing in NYSUT's total. It appears that retaining both would double-count a fifth of New York State's unionized teachers. The same logic applies to locals that genuinely represent workers across a wide geography from a single office, e.g., SEIU~1199, which spans six states and the District of Columbia, or SEIU~2015, which represents long-term-care workers throughout California. Attributing 1199's 400{,}000 members to a single New York address, or all of 2015's unionized home-care workforce to Los Angeles County injects significant error into exactly the local geographies we are trying to validate.

The shared-address rule targets a different situation: unions organized by employer or craft rather than geography. For example, local unions of the Air Line Pilots Association are organized by airline. All ALPA locals file LM forms from a single address in McLean, Virginia.  Similarly the AFA-CWA flight-attendant locals all file from an address in Washington, D.C. These are not local memberships in any geographic sense, and a shared address is a signal of the pattern. Because the top-one-percent and shared-address rules are conceptually distinct---one flags implausible size, the other implausible geographic concentration---the exclude version applies both, and we report our validation results against both versions.  We do not argue for the superiority of one over the other but rather intend to show the degree of sensitivity to reasonable data cleaning and attribution choices.

\section{The validation pipeline}

This section outlines the logic of our validation pipeline and summarizes the results that we develop in detail in subsequent sections.  To circumvent the lack of direct ``ground truth'' for substate unionization,  we rely on a model-based triangulation scheme.  The logic is as follows: we link well-understood CPS-based estimates of union density with less certain LM-based estimates via multilevel modeling with post-stratification (MrP) \citep{gelmanPoststratificationManyCategories1997, parkBayesianMultilevelEstimation2004, ghitzaDeepInteractionsMRP2013}.   MrP is a widely-used statistical technique for small area estimation, i.e., combining national-level survey data with population counts to generate reliable, regularized subnational estimates. That is, we use the CPS to train individual-level models of union membership in 2000, 2007, 2014, and 2019.\footnote{To stabilize estimates, we follow \citet{cardCanFallingSupply2001} and pool consecutive years of the CPS; i.e., 2000 is trained on the 1999--2001 CPS. } Then, using the 2000 Census and the 2007, 2014, and 2019 American Community Survey (ACS), we construct post-stratification tables for the joint frequency distribution of all the predictors in the individual model at the state, 1990 CZ, CD, and county levels.\footnote{The ACS is a far larger survey.  For 2019, the ACS 5-year extract we use had 6.4 million respondents compared to 448,000 for the 3-year pooled CPS.}  Applying the trained individual-level model to the post-stratification tables produces estimates of union membership at these differing geographies.  As a modeling strategy, MrP is designed to ``share strength'' across units to produce more accurate and less variable estimates in sparsely-populated cells.  There is precedent for using MrP to generate subnational unionization estimates from survey data; \citet{agnolinRobotsReplacingTradeforthcoming} produce just such a dataset for several European countries and the NUTS2 level.  

We look at counties, CDs, and CZs for several reasons.  On a practical basis, these are the geographies to which scholars have attempted to aggregate LM-based reports of union membership.  ACS survey data are available or can be mapped to these geographies for the construction of poststratification tables.  These geographies also represent different sorts of aggregation; it is conceivable that LM-based aggregation might work better for certain geographies than others.  CDs are designed to have very similar population levels and they are contained within state boundaries, but their shapes and demographics vary endogenously over time.  CD boundaries do not necessarily respect other substate political units or regional economies.  Commuting zones, on the other hand, are meant to represent local labor markets. They tessellate the entire US and they are widely known as the unit of analysis in much of the ``China shock'' literature.  We use fixed 1990 commuting zones for consistency and stability.  But CZs vary substantially in size and employment structure and they can cross state boundaries.  Finally, counties are relatively stable over time and represent important substate political units, but they vary widely in size and population and the ACS only provides data sufficient for post-stratification for the 475 largest, most populous US counties.  We are therefore unable to aggregate county-level estimates to the state level. 

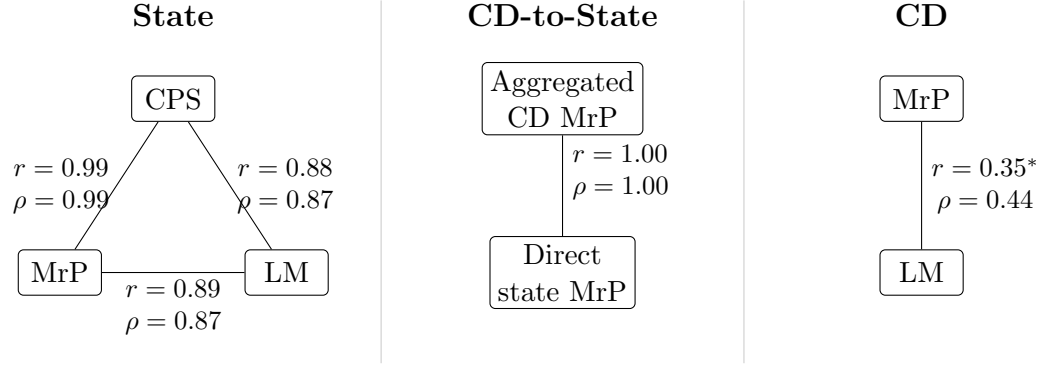
\begin{figure}\caption{The validation logic, from CPS-based union density at the state level to MrP and LM-based density at the Congressional district level.  Correlation coefficients represent averages across those from all four years. For edges involving LM, the value shown is the higher of the inclusive and exclusive membership definitions; an asterisk ($^{*}$) marks a value drawn from the exclusive definition.}\label{fig:schematic_summary}
\begin{tikzpicture}[
    src/.style  = {draw, rounded corners=2pt, align=center, fill=white,
                   inner sep=3pt, minimum width=1.1cm, minimum height=0.6cm,
                   font=\small},
    ttl/.style  = {font=\bfseries},
    elab/.style = {align=center, font=\footnotesize, inner sep=1.5pt}]

  \draw[black!20] (4.75,-0.1) -- (4.75,4.7);
  \draw[black!20] (9.55,-0.1) -- (9.55,4.7);

  \node[ttl] at (2.0,4.5) {State};
  \node[src] (p1cps) at (2.0,3.4) {CPS};
  \node[src] (p1mrp) at (0.5,1.1) {MrP};
  \node[src] (p1lm)  at (3.5,1.1) {LM};

  \draw (p1cps) -- (p1mrp)
        node[elab, midway, left=1pt] {$r=0.99$\\$\rho=0.99$};
  \draw (p1cps) -- (p1lm)
        node[elab, midway, right=1pt] {$r=0.88$\\$\rho=0.87$};
  \draw (p1mrp) -- (p1lm)
        node[elab, midway, below=1pt] {$r=0.89$\\$\rho=0.87$};

  \node[ttl] at (7.15,4.5) {CD-to-State};
  \node[src] (p2agg) at (7.15,3.4) {Aggregated\\CD MrP};
  \node[src] (p2dir) at (7.15,1.1) {Direct\\state MrP};

  \draw (p2agg) -- (p2dir)
        node[elab, midway, right=2pt, align=left]
        {$r=1.00$\\$\rho=1.00$\\};

  \node[ttl] at (11.9,4.5) {CD};
  \node[src] (p3mrp) at (11.9,3.4) {MrP};
  \node[src] (p3lm)  at (11.9,1.1) {LM};

  \draw (p3mrp) -- (p3lm)
        node[elab, midway, right=2pt] {$r=0.35^{*}$\\$\rho=0.44$};

\end{tikzpicture}
\end{figure}

Figure~\ref{fig:schematic_summary} lays out the pipeline in schematic form and summarizes findings by labeling the edges with Pearson's $r$ and Spearman's $\rho$ reflecting the level of correspondence between union density estimates from each data source.  In the figure we use CDs as our substate unit as CD-to-state aggregation is straightforward.  In the left panel, we see that our model-based state-level MrP estimates generate near perfect agreement with the CPS ($r, \rho\geq 0.99$), as one would expect.  We reproduce \citet{becherLocalUnionOrganization2018b}'s  weaker but still impressive state-level agreement between LM-based estimates and both CPS and MrP estimates ($r, \rho\geq 0.87$).  In the middle panel, we show that MrP estimates aggregate well: post-stratifying at the CD level and then aggregating unionization estimates up to the state-level produces estimates that correlate strongly ($r, \rho>0.99$) with estimates produced by post-stratifying directly at the state-level.  This tight CPS-MrP correspondence is the conceptual anchor for the triangulation strategy, licensing MrP estimates as the stand-in CPS at lower levels.  

After establishing that (i) MrP estimates generate highly accurate state-level estimates and (ii) CD-level MrP estimates aggregate to the state level, we also show that MrP and CPS estimates correspond closely among the few counties with sufficient data for reliable comparison.  Along the way we show that MrP estimates also generate reasonable within-state variation in union membership.\footnote{Demonstrating within-state variation is important, as the individual-level MrP models that carry state effects might simply ``memorize'' the state-level averages from the training data.  We show that this did not occur with the model we selected, although within-state variation in MrP estimates is lower than CPS due to ``partial pooling'' in hierarchical models.}

Having demonstrated the applicability of substate MrP estimates, we use them as the basis for calibrating LM-based estimates of unionization at finer geographies.  Even if MrP and LM-based estimates are imperfect, reliable LM-based estimates should exhibit the same strong correlation with MrP estimates at the substate level that they show at the state level.  Unfortunately, this is not what we find.  Not only does the LM-MrP correspondence break down dramatically at the CD, CZ, and county levels (Figure~\ref{fig:descent}), but we also find significant lumpiness in the LM data that yields union density levels $>1.0$ at the county and CD levels. Moreover, the severity of the pathologies depends on data cleaning choices designed to address lumpiness in LM membership reporting.

\subsection{Building a MrP model for unionization}\label{sec:MrP}
\subsubsection{Model evaluation and selection}
\label{sec:model-selection}

Our validation exercise requires constructing a predictive model for union membership at the individual level that nevertheless aggregates accurately at different geographic scales. We adopt a machine learning approach, constrained by the need to have comparable data in both the estimation (CPS) and post-stratification (ACS) datasets while avoiding sparse demographic cells.\footnote{That is, the number of post-stratification cells grows exponentially in the number of strata.}  We specified multiple different individual-level models and evaluated out-of-sample performance. In an MrP framework, better-performing predictive models at the individual level may not be the best at the level of post-stratified aggregation \citep{kuhUsingLeaveoneoutCross2024, kennedyModelValidationAggregate2024}.  We therefore examine model predictive performance at the individual level and at the state level.

We considered modeling approaches ranging from simple logistic regression to gradient-boosted trees; details appear in Appendix~\ref{app:model-selection}.  All models were trained on pooled monthly CPS outgoing-rotation-group (ORG) samples \citep{ellisIpumsrInterfaceDownloading2026, flood2025ipumscps} from three-year windows centered on 2007 and 2019 (2006--2008 and 2018--2020).\footnote{Model comparison used two of the four analysis years. The final fits cover 2000, 2007, 2014, and 2019.} The richer models share the same individual predictors that coincide with both the CPS and the ACS poststratification tables---age (4 categories), race/ethnicity (4), sex (2), education (5), and industry/sector (5).\footnote{We combine industry and sector to highlight those most associated with unionization while minimizing the growth in the number of strata.  Each CPS or ACS respondent is assigned to one of five mutually exclusive industry-sector categories: federal, state, or local government workers are all coded public; everyone else is classified from their 1990-basis industry code into \{construction/extraction/mining, manufacturing, transportation/communications/utilities, other\}. We also explored occupational categories, but preliminary analysis suggested the occupation and industry/sector picked up the same variation.}   The tree ensembles were tuned by three-fold cross-validation on the training split, while the hierarchical Bayesian models used weakly informative priors and fit using MCMC, via \texttt{rstanarm} \citep{goodrichRstanarmBayesianApplied2025, standevelopmentteamRStanInterfaceStan2025}.

For model evaluation and selection purposes we fit the full train/test/validate pipeline separately for each comparison year. From each year's pooled sample we drew an 80/20 train/test split stratified on union membership. We evaluated models on the individual-level discrimination on the held-out test split and the recovery of the state-level aggregate, comparing each model's post-stratified state density estimate against the CPS.  Among models performing similarly, we then looked at the variation of post-stratified density across congressional districts within each state. As estimating within-state variation is the reason for turning to LM data in the first place, we prefer models producing greater variation within-states, holding fixed their predictive accuracy.

Among the candidate models, we selected a Bayesian hierarchical logit\footnote{Full specifications, tuning, and per-model performance appear in Appendix~\ref{app:model-selection}.   Full details for the selected model are in Appendix~\ref{app:shl_selected}.}. Although training the hierarchical model is more computationally expensive relative to XGBoost, its aggregate performance is substantially better and it produces a full posterior predictive distribution, yielding a credible interval for every area estimate that the ensemble methods do not provide.  More complex models including fully-saturated demographic $\times$ state interactions provided no appreciable benefit but imposed a higher computational cost. 

Having selected the hierarchical logit, we train the model separately on CPS ORG data for each year in \{2000, 2007, 2014, 2019\} via \texttt{rstanarm}. Each training run consisted of four independent Markov chains of 2000 iterations each, discarding the first 1000 as burn-in. Convergence is assessed by the potential scale reduction factor $\widehat{R}$ and the effective sample size: the maximum $\widehat{R}$ is at most 1.004 and the minimum effective sample size is at least 876, suggesting all the chains converged to stationary posterior distributions (Appendix Table~\ref{tab:mrp-convergence}).

\subsection{Post-stratification at the state, county, commuting zone, and Congressional district levels}
\label{sec:poststrat}

We generate density estimates at each geography by post-stratifying the fitted individual-level model over population cells constructed from the American Community Survey (ACS) and 2000 US census.\footnote{For the ACS, we use the 5-year extract centered on the year in question to maximize the number of observations in potentially sparse demographic cells.  We use a 5 percent sample for the 2000 census.} Post-stratification cells are defined by the full joint frequency table of the model's demographic factors---age, sex, education, race, industry/sector---within each geographic unit.\footnote{The final poststratification table had 800 cells per geographic unit.} For each cell we take the model's mean posterior predicted probability of union membership and aggregate these predictions to the unit level, weighting each cell by its population. We carry the full posterior distribution through this aggregation---computing a density estimate for every posterior draw and only then summarizing---rather than collapsing to point predictions beforehand.  The denominator for union density is the ACS-reported employed wage-and-salary earners over 16, corresponding to the definition the CPS uses for calculating union density. 

The geographic resolution at which we can post-stratify is limited by available data and geographic identifiers in the public-use ACS microdata. Specifically, ACS builds up from PUMAs.  It only identifies counties when doing so does not pose a privacy concern.  In practice this means that about 475 out of approximately 3100 US counties are identified with the full demographic frequency distribution.\footnote{These are mostly the largest counties but some large counties are not identified in the ACS.}  We restrict county-level analysis to this identified subset, which covers a majority of the US employed population. Congressional districts and 1990 commuting zones are not identified in the microdata; we construct them by allocating each PUMA's population to CDs and CZs through PUMA-to-CD crosswalk from the US Census Bureau and the  PUMA-to-CZ crosswalks from David Dorn \citep{autorGrowthLowSkillService2013, autorWhenWorkDisappears2019}. Each five-year ACS extract is paired with the crosswalk matching its PUMA vintage. We obtain MrP estimates for all 435 congressional districts and all 741 1990 commuting zones in each year.

Commuting zones introduce one complication that congressional districts do not: CZs can cross state lines, whereas the model carries a state-specific effect. CDs nest cleanly within states, so each district simply inherits its state's effect. For CZs, we instead form the state effect as an employment-weighted blend of the relevant states' effects, applied separately within each posterior draw: for a zone spanning several states we combine those states' draw-level effects in proportion to the zone's employment in each.\footnote{In a small number of cases a PUMA fails to match or falls outside any district, producing minor shortfalls between district-summed and state-total employment; this is negligible except in Louisiana in 2007, where district-summed employment falls roughly four percent short.  We exclude 2007 Louisiana from analyses that aggregate districts to the state level.}

\subsection{Reliability of MrP estimates}

To establish the reliability of MrP substate estimates before using them as a benchmark we undertake two tests.  First, we ask whether MrP substate estimates aggregate well to the state level, where MrP's performance is known.  That is, if we generate CD-level estimates of union density and then aggregate those to the state level, do we recover state-level estimates close to what we calculate if we generate MrP estimates at the state level directly?  We use CDs as they are completely within-state and they cover all states in their entirety.

\begin{figure}[htbp]
  \centering
  \includegraphics[width=\linewidth]{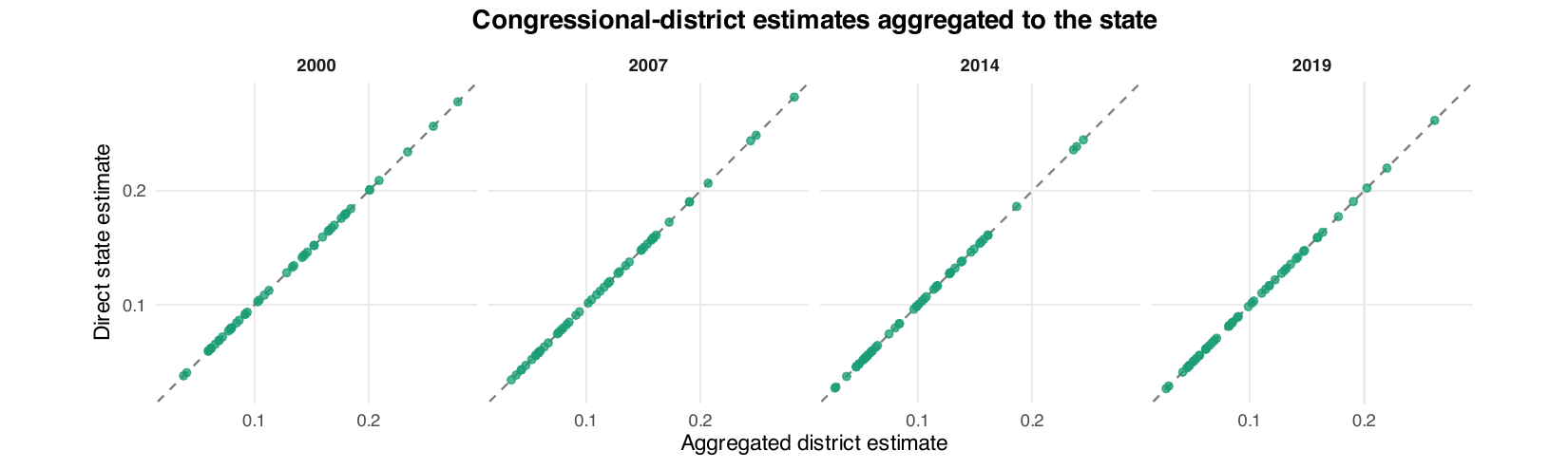}
  \caption{Congressional district-level MrP estimates of union density to the state level produce state-level estimates that are strongly correlated with MrP state-level estimates.}
  \label{fig:mrP_agg_cd_to_state}
\end{figure}

Figure~\ref{fig:mrP_agg_cd_to_state} plots state-level MrP density estimates against CD-level density estimates aggregated up to the state level (weighted by CD share of state employment). CD-level MrP does, in fact, aggregate accurately, generating state-level density estimates that are tightly linked with direct MrP state level estimates. 

\begin{figure}[htbp]
    \centering
    \includegraphics[width=0.7\linewidth]{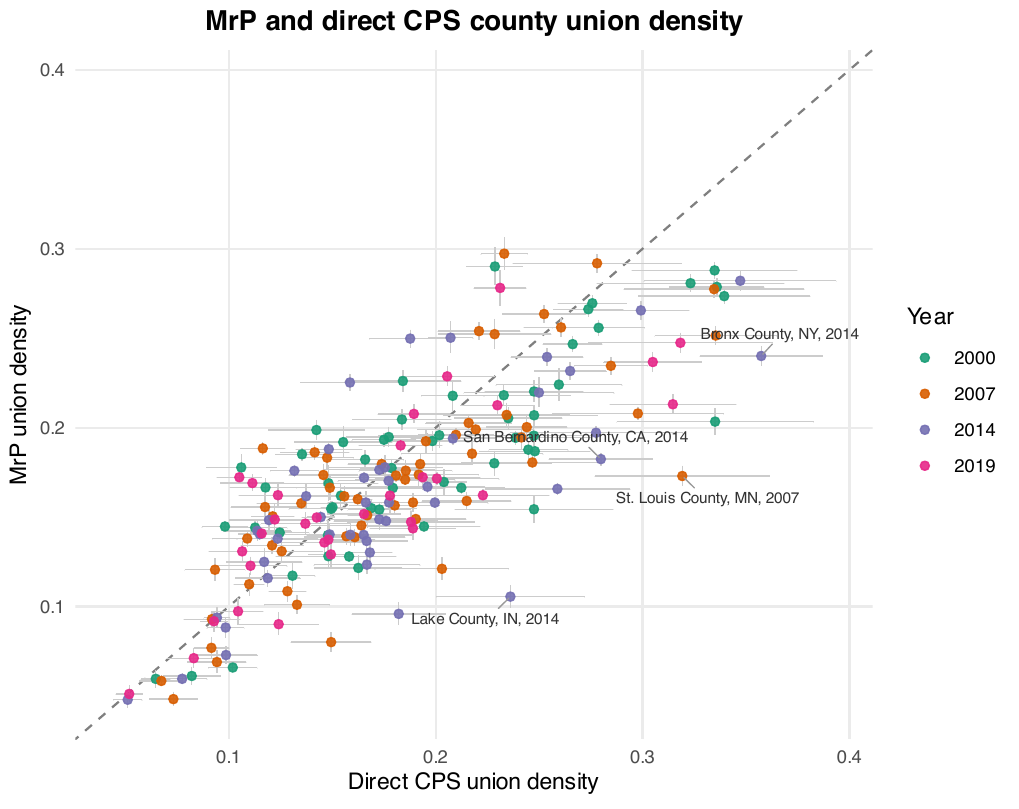}
    \caption{MrP union density estimates versus direct CPS union density for 192 county-years in 82 counties (2000, 2007, 2014, 2019). Each point is one county-year. Horizontal bars are the CPS 95\% sampling interval, vertical bars the MrP 95\% credible interval around the posterior mean.}
    \label{fig:mrp-vs-cps-county}
  \end{figure}

Second, we directly compare a limited number of county-level CPS estimates to those generated by MrP.  Specifically, we examine the counties with sufficient CPS data to provide a plausible estimate of union density in a given year.  This generates 192 county-years from 82 counties.\footnote{From the counties with available data for post-stratification, we keep a county in a target year if it is identified in the CPS-ORG in all three years of that year's CPS window and its pooled CPS sample over the window contains at least 125 union members; a county can qualify in some target years and not others.}  We compare these with the MrP density estimates for the same county-years.  Results appear in Figure~\ref{fig:mrp-vs-cps-county}.

We see a strong correspondence ($r = 0.80$, $\rho = 0.79$) between the county-level MrP estimates and those available from the CPS.  The error bars give some indication of the uncertainty around these estimates. Overall, 77 of 192 county-years (40\%) have MrP posterior means within the 95\% sampling interval of the CPS estimates.  This is far better performance than either version of the LM data for these same counties (Appendix~\ref{app:CPS_county_v_LM}, Figure~\ref{fig:lm-vs-cps-county}).  MrP estimates appear as a reliable benchmark for the CPS at substate geographies.

\subsubsection{Within-state variation in unionization}

Showing that MrP can extract meaningful within-state variation is critical to justifying its use as a reasonable benchmark for substate quantities.  The county-level analysis allows for just this.

\begin{figure}[htbp]
    \centering
    \includegraphics[width=0.7\linewidth]{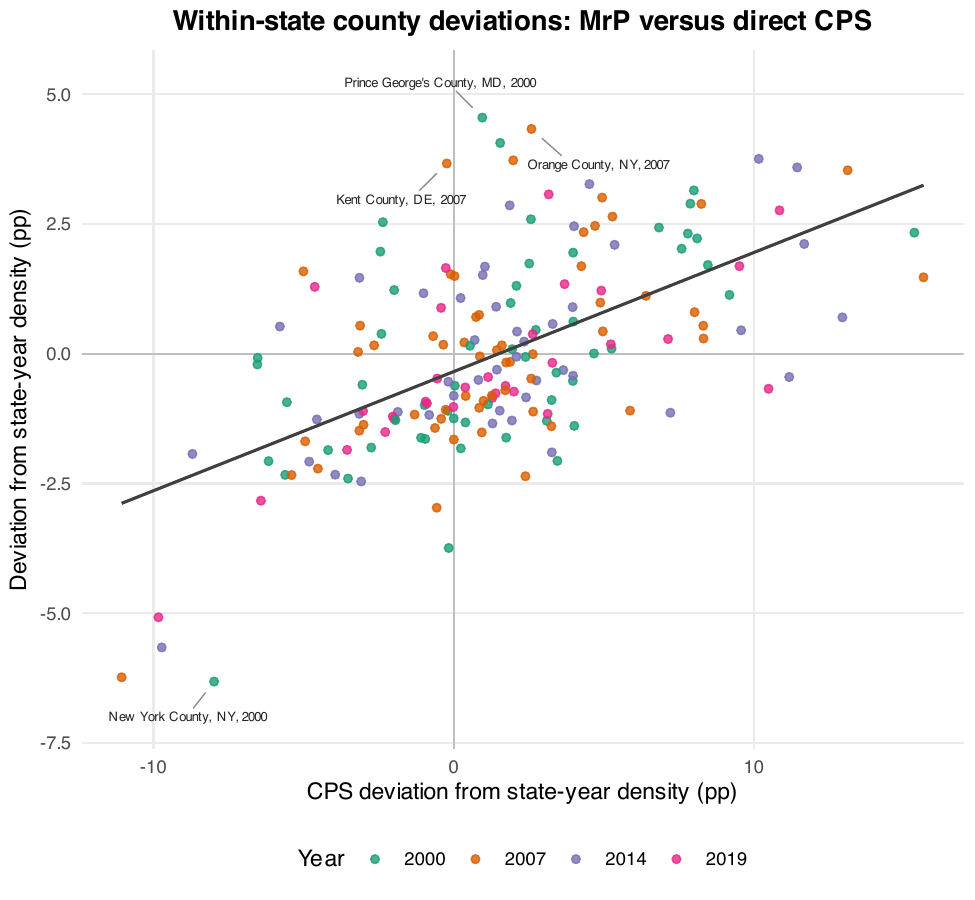}
    \caption{MrP within-state union density deviations versus CPS within-state deviations for 188 county-years in 81 counties (2000, 2007, 2014, 2019). Each county-year is expressed as a deviation from its own state-year density in the same measure. DC is excluded.}
    \label{fig:mrp-lm-cps-county-deviation}
  \end{figure}

Figure~\ref{fig:mrp-lm-cps-county-deviation} plots MrP county-level deviations from their state-year densities against the same quantity produced using CPS across 188 county-years in 81 counties (we drop DC here).  The MrP estimates have a mean absolute deviation from their state-year average of 1.4 percentage points, roughly 39\% of the mean absolute within-state deviation for the CPS (3.7pp).  MrP, as a shrinkage estimator, appears to understate within-state variation relative to the CPS while still recovering meaningful within-state differences in unionization rates. Importantly, the within-state deviations between MrP and CPS correlate at about 0.6.\footnote{Appendix Table~\ref{tab:within_state_county_validation} reports the within-state correlations and mean absolute deviations for MrP and both LM definitions on the identical sample.}  This contrasts sharply with the LM-based estimates, where within-state deviations are virtually uncorrelated with those from CPS.\footnote{The corresponding plot displaying LM against CPS appears in Appendix Figure~\ref{fig:lm-cps-county-deviation}.}      

\section{Results: agreement at the state level but divergence below}
\label{sec:ladder}

This section presents the detailed findings of the full validation pipeline.  In making comparisons we use both versions of the LM dataset: the ``exclusive'' version (omitting all union locals that share an address as well as the top 1\% by membership) and the ``inclusive'' version that keeps those locals in the data set. 

\subsection{State Level: Establishing the Anchor}
\label{sec:ladder-state}

\begin{figure}[htbp]
  \centering
  \includegraphics[width=\linewidth]{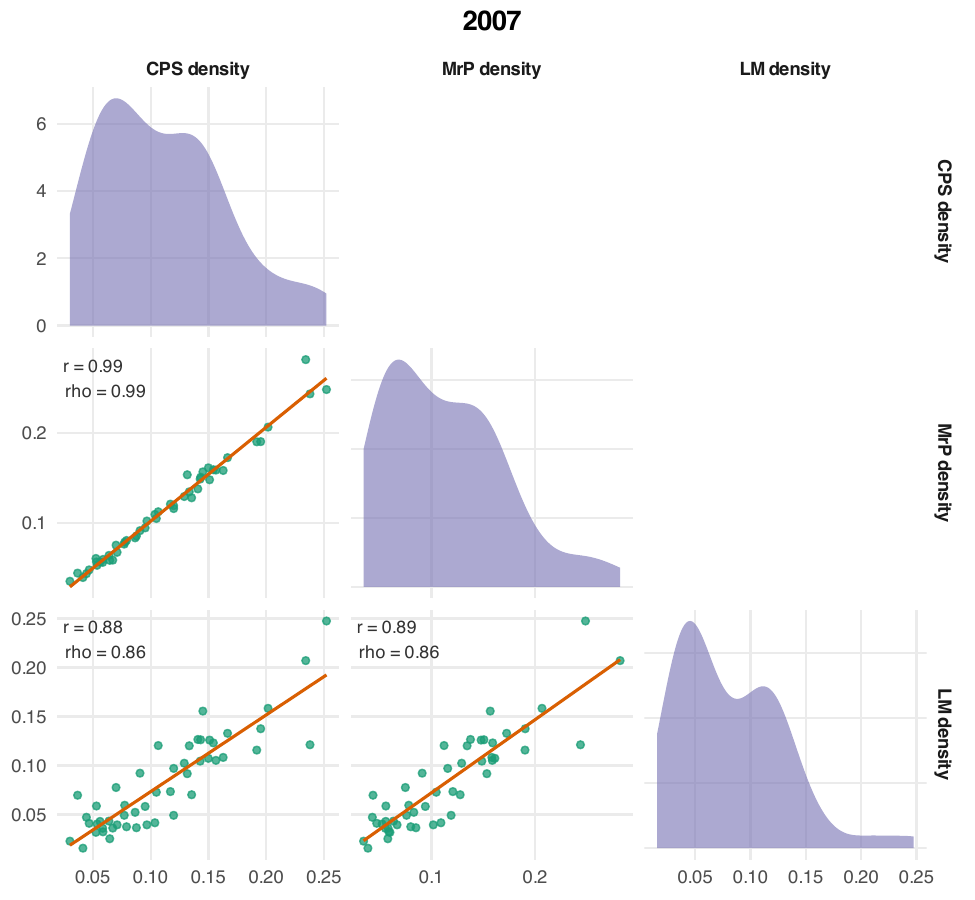}
  \caption{Union density at the state level, 2007: pairwise comparison of CPS, MrP, and LM, using the inclusive LM membership definition.}
  \label{fig:state_matrix_density_include_2007}
\end{figure}

We first establish that the CPS (as reported in Unionstats), MrP, and LM-derived estimates of union density correlate highly at the state level.  Figure~\ref{fig:state_matrix_density_include_2007} displays the univariate distributions as well as pairwise state-level correlations for 2007, using the inclusive LM-derived membership definition.  We see very strong correlations, both in levels and ranks between CPS and MrP.  The LM-CPS and LM-MrP correspondence is weaker but still quite strong.  Results for the other years are similar and are collected in Appendix~\ref{app:complete_figs} (Figures~\ref{fig:state_matrix_density_include_2000}--\ref{fig:state_facet_density_include_mrp_lm}).    

In the appendix we report these relationships using the exclusive version of the LM data (Figures~\ref{fig:state_matrix_density_exclude_2000}--\ref{fig:state_facet_density_exclude_mrp_lm}).  As expected, the exclusive version produces weaker correlations than the inclusive version with both CPS (2007 $r=0.7, \rho=0.8$) and MrP (2007 $r=0.6, \rho=0.7$) at the state level.

\subsection{Descending the hierarchy}
\label{sec:ladder-descent}

\begin{figure}[htbp]
  \centering
  \includegraphics[width=\linewidth]{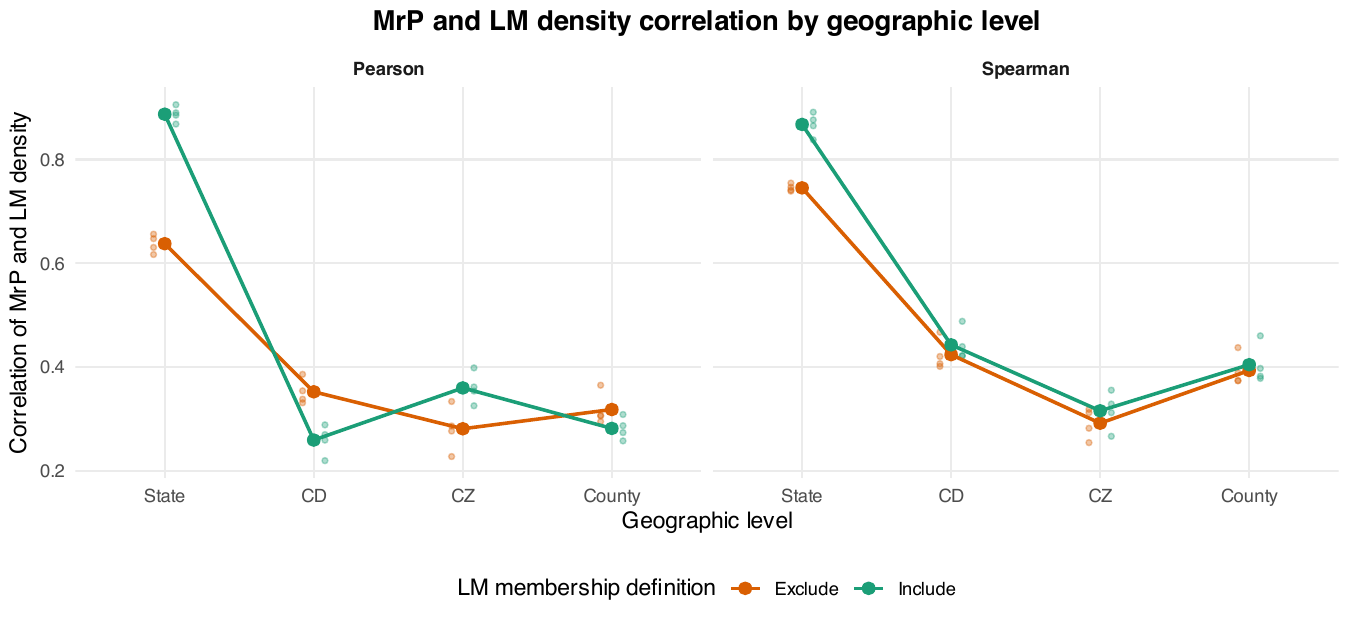}
  \caption{Correlation between MrP and LM union density by geographic level, for the inclusive and exclusive LM membership definitions, shown separately for Pearson (left) and Spearman (right) correlation. The small points are the four individual per-year correlations (2000, 2007, 2014, and 2019), each computed within that year; the large connected points are the unweighted average of those four per-year correlations. DC and Louisiana~2007 are excluded at the state, congressional-district, and county levels; all 741 commuting zones are retained.}
  \label{fig:descent}
\end{figure}

As we go to finer levels of geographic resolution, the correspondence between LM and MrP estimates of density weakens dramatically.  Figure~\ref{fig:descent} summarizes these relationships at each geographic level, for both definitions of LM membership and both correlation metrics.  At the substate level, the MrP-LM correlation is less than half of what it is at the state level.  Pearson correlation hovers around a modest 0.3, regardless of the LM measurement decision, ranging from 0.40 in 2000 (CZ, inclusive) to 0.22 in 2019 (CD, inclusive).  There is no obvious pattern here over time or geographic scale, although 2019 produced the weakest correspondences across all years and levels considered.  This is not an artifact of the correlation metric, as we see quite similar patterns in the Spearman correlation in density rankings.   

\subsection{Exploring the sub-state divergence between MrP and LM}

\begin{figure}[htbp]
  \centering
  \begin{subfigure}{\linewidth}
    \centering
    \includegraphics[width=\linewidth]{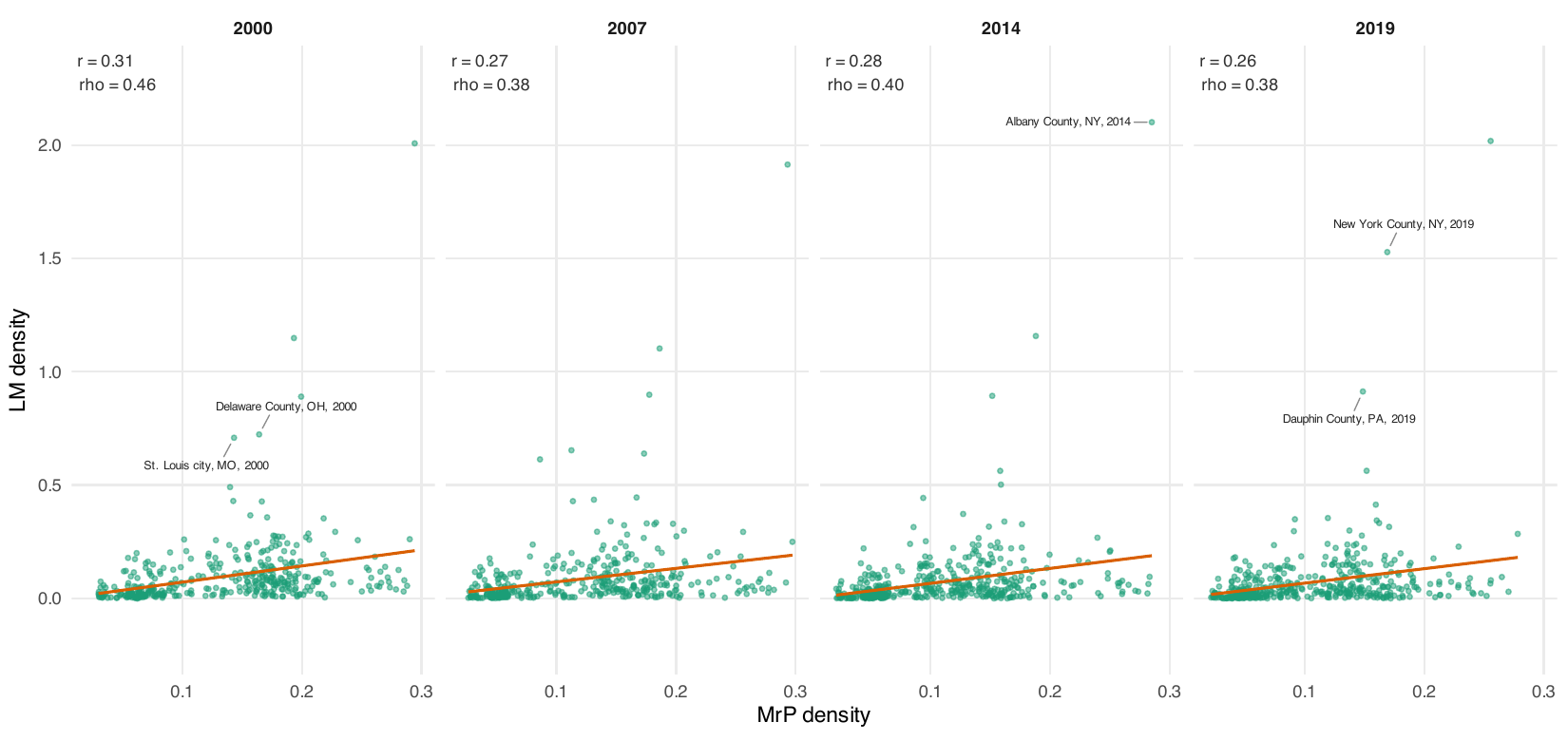}
    \caption{Inclusive LM membership definition.}
    \label{fig:county_facet_mrp_lm_density_include}
  \end{subfigure}

  \vspace{1em}

  \begin{subfigure}{\linewidth}
    \centering
    \includegraphics[width=\linewidth]{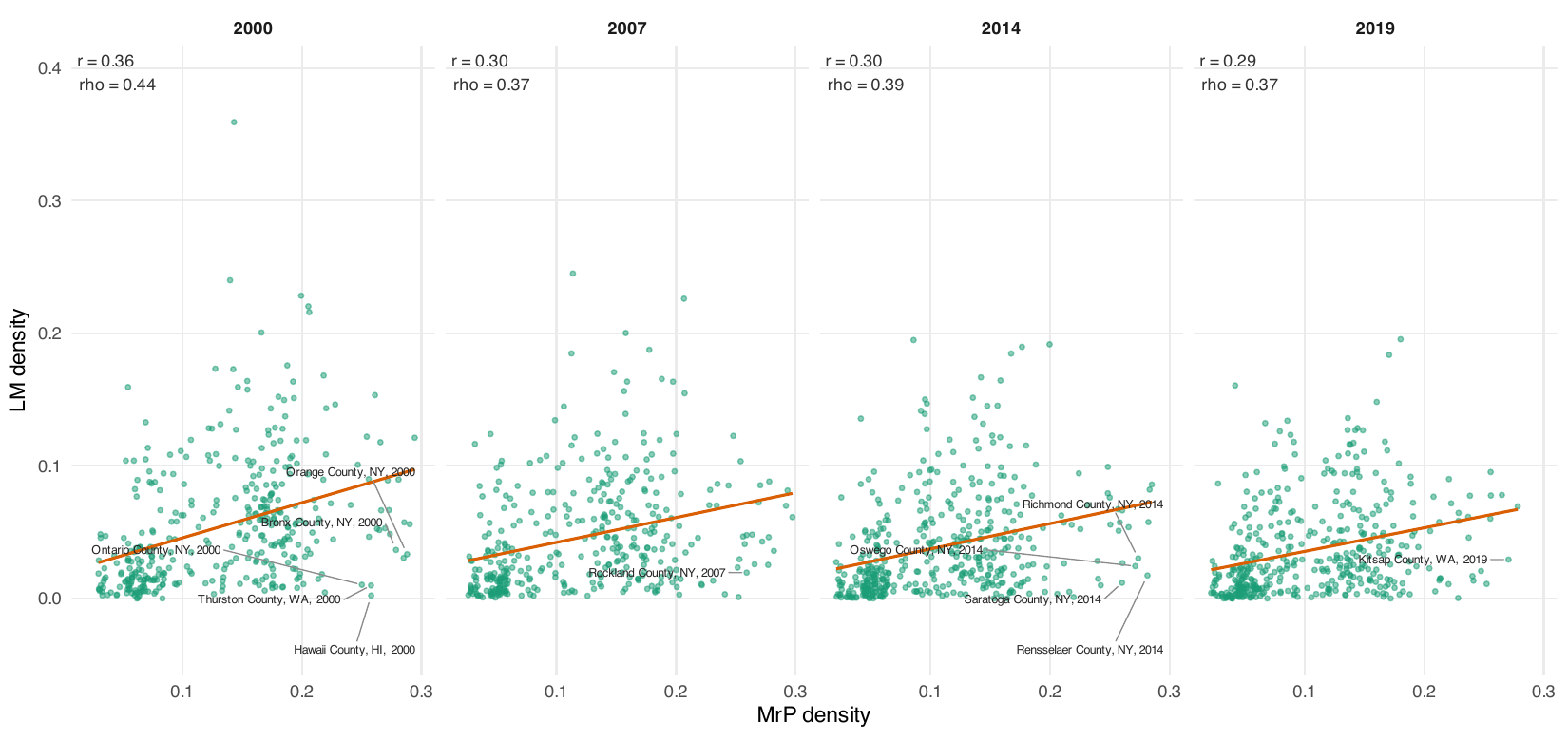}
    \caption{Exclusive LM membership definition.}
    \label{fig:county_facet_mrp_lm_density_exclude}
  \end{subfigure}

  \caption{Union density at the county level: MrP versus LM, faceted by year
    (2000, 2007, 2014, 2019). Panel (a) uses the inclusive LM membership
    definition; panel (b) the exclusive definition.}
  \label{fig:county_facet_mrp_lm_density}
\end{figure}

Figure~\ref{fig:county_facet_mrp_lm_density} displays the MrP-LM correspondence at the county level.\footnote{Appendix~\ref{app:complete_figs} displays the same figures at the CD and CZ levels.}  Counties for which we have ACS data are largely fixed over our time frame.  They are fully contained within state boundaries but they vary in size tremendously. Comparing panels \ref{fig:county_facet_mrp_lm_density_include} (inclusive LM) and \ref{fig:county_facet_mrp_lm_density_exclude} (exclusive), we see two things.  First, at the county level, the exclusive version of LM correlates with MrP marginally better than the inclusive version, the opposite of the state level pattern.  Examining the vertical axis in both panels indicates why: lumpiness in reporting.  In the inclusive version of the LM data, there are several counties showing union densities near or in excess of 1.0.  That is, county-aggregated union membership as reported on LM forms is larger than the county's entire employed population. While this is conceptually possible---for example, if union locals were to report large numbers of retired and/or unemployed members on LM forms linked to an otherwise small county---it is very unlikely.  Rather, it appears to be the result of many unions having mailing addresses in moderate population counties, often because it is a state capital or former industrial zone in a state with a history of unionization.  For example, the largest density value (210\%) is for Albany county, NY, which includes the NYSUT.\footnote{The largest observed county-level union density from MrP estimates is 0.30 (Honolulu County, HI, 2007).}  Other noteworthy examples include Dauphin County, PA (Harrisburg area) and St. Louis City, which is the small former industrial core that is surrounded by the much more populous (and separate) St. Louis County.  This pathology disappears when using the exclusive version of the LM data that omits the extremely large locals and those that share addresses.  This suggests that lumpiness in reporting that fails to correspond to worker residential patterns is the culprit here.

Table~\ref{tab:lm-impossible} drills into this further, reporting the number of geographic units whose LM-based union density exceeds 1.0. Using the inclusive definition, some congressional districts and counties cross this threshold, while under the exclusive definition no unit exceeds 1.0 at any geography.  Moreover, the maximum union density recorded under the inclusive definition is unreasonably large.  The exclusive version of the LM data avoids these problems at the extremes.  In other words, the substate estimates of union density using LM data display significant lumpiness, even producing implausibly large values. The forces producing this lumpiness--extremely large locals and those sharing addresses--are the result of \emph{organizational} factors and decisions (e.g., how union locals will be structured), not the result of the geographic dispersal of members around some central location.     

\begin{table}[htbp]
\centering
\caption{Number and maximum of geographic units whose LM-aggregated union density exceeds 1.0 (more reported members than employed workers), by LM membership definition, pooled over 2000, 2007, 2014, and 2019. The first two value columns use the inclusive definition, the last two the exclusive definition.}
\label{tab:lm-impossible}
\begin{tabular}{lr rr rr}
\toprule
 &  & \multicolumn{2}{c}{Inclusive} & \multicolumn{2}{c}{Exclusive} \\
\cmidrule(lr){3-4}\cmidrule(lr){5-6}
Geography & Units & Number $>1$ & Maximum & Number $>1$ & Maximum \\
\midrule
State & 204 & 0 & 0.65 & 0 & 0.24 \\
CD & 1744 & 13 & 2.63 & 0 & 0.35 \\
CZ & 2964 & 0 & 0.63 & 0 & 0.28 \\
County & 1613 & 8 & 2.10 & 0 & 0.36 \\
\bottomrule
\end{tabular}

\end{table}

As another illustration of how aggregating LM reports to smaller geographies generates lumpiness, we look at the signed gap between MrP and LM-reported union density at the CD level.  Table~\ref{tab:gap_variance_decomposition} decomposes the standard deviation of this density gap into between-state and within-state components, by year and LM definition. Under the inclusive definition within-state variation dominates. Variation across CDs in the same state accounts for about 95\% of the total variation in the density gap in every year.  On average, LM estimates differ from MrP by around 11 percentage points within a state (a huge amount). Under the exclusive definition, the within-state variation in the MrP-LM gap falls sharply---the typical district now differs from others in the same state by around three to four percentage points.  Across-state differences now account for the majority of the gap.  These two versions of the LM data paint very different pictures about the location and sources of variation in union membership in the US.\footnote{Appendix Figure~\ref{fig:district_gap_strip} shows the individual district gaps within each state relative to that state's density gap for the exclusive definition (Appendix~\ref{app:district-gap-strip}); Appendix~\ref{app:district-structure} gives the inclusive analog, the pooled district-gap distribution, and the district-level public-sector regressions.}

\begin{table}[htbp]
\centering
\caption{Decomposition of the congressional district-level MrP-LM density gap into between-state and within-state components, by year and LM definition. Between-state component is the standard deviation of the state mean district gap, and the within-state component is the square root of the mean, across states, of the within-state variance (single-district states excluded). DC and Louisiana~2007 excluded.}
\label{tab:gap_variance_decomposition}
\begin{tabular}{l rrr rrr}
\toprule
 & \multicolumn{3}{c}{Inclusive} & \multicolumn{3}{c}{Exclusive} \\
\cmidrule(lr){2-4}\cmidrule(lr){5-7}
Year & Between (pp) & Within (pp) & Within share & Between (pp) & Within (pp) & Within share \\
\midrule
2000 & 2.47 & 11.17 & 0.95 & 4.55 & 4.04 & 0.44 \\
2007 & 2.67 & 11.44 & 0.95 & 4.82 & 3.57 & 0.35 \\
2014 & 2.47 & 10.69 & 0.95 & 4.50 & 3.17 & 0.33 \\
2019 & 2.59 & 12.23 & 0.96 & 4.43 & 2.98 & 0.31 \\
\bottomrule
\end{tabular}

\end{table}

All of this shows up starkly when we consider the spatial correlation in union density. Table~\ref{tab:moran-2014} reports the global Moran's $I$ for each of the density measures at the CD and CZ level for 2014.\footnote{Moran's $I$ is a within-dataset measure that summarizes how similar close-by units are to one another. We used a row-standardized ``queen'' contiguity matrix. DC is included here but zero-neighbour isolates are excluded.}  The MrP benchmark shows a high degree of spatial correlation: CZs and CDs near each other tend to have similar levels of local union density.  But spatial correlation collapses for the LM-include values, reflecting the reporting lumpiness we just described.  The exclude version of the LM-based density estimates show somewhat more spatial similarity, but nowhere near what we see in the MrP benchmark. 

\begin{table}[htbp]
\centering
\caption{Global Moran's~$I$ of 2014 union density at the CZ and CD levels, computed separately for MrP, inclusive LM, and exclusive LM. }
  \label{tab:moran-2014}
  \begin{tabular}{ll r r}
\toprule
Geography & Measure & Moran's $I$ & $N$ \\
\midrule
Commuting zone & MrP density & 0.855 & 734 \\
Commuting zone & LM density (include) & 0.204 & 734 \\
Commuting zone & LM density (exclude) & 0.311 & 734 \\
Congressional district & MrP density & 0.873 & 435 \\
Congressional district & LM density (include) & 0.158 & 435 \\
Congressional district & LM density (exclude) & 0.239 & 435 \\
\midrule
\multicolumn{4}{@{}l@{}}{\footnotesize Zero-neighbour isolates excluded.} \\
\bottomrule
\end{tabular}

  \end{table}

Table~\ref{tab:lee-2014} presents Lee's $L$, a measure of the spatial similarity across datasets that answers whether two measures agree on \emph{where} unionization is high or low.\footnote{$L$ for two variables $x$ and $y$ ranges in $[-1,1]$.  One way of expressing $L$ is as the product of the spatial autocorrelation in $x$ with the same for $y$ multiplied by the Pearson correlation $r(x,y)$.}  We see that the sub-state correlation between the MrP benchmark and the LM-derived measures is even weaker, i.e., they diverge in the spatial patterns of union density that they generate.  Worryingly, the correlation \emph{between} the inclusive and exclusive versions of the density measure is equally poor or even worse, suggesting that the lumpiness in LM reporting is extreme.

  \begin{table}[htbp]
  \centering
  \caption{Lee's~$L$ spatial cross-correlation between pairs of 2014 union density measures, at the CZ and CD levels. DC is included but zero-neighbour islands are excluded.}
  \label{tab:lee-2014}
  \begin{tabular}{ll r r}
\toprule
Geography & Measure pair & Lee's $L$ & $N$ \\
\midrule
Commuting zone & MrP vs LM (include) & 0.319 & 734 \\
Commuting zone & MrP vs LM (exclude) & 0.255 & 734 \\
Commuting zone & LM include vs LM exclude & 0.303 & 734 \\
Congressional district & MrP vs LM (include) & 0.276 & 435 \\
Congressional district & MrP vs LM (exclude) & 0.314 & 435 \\
Congressional district & LM include vs LM exclude & 0.196 & 435 \\
\midrule
\multicolumn{4}{@{}l@{}}{\footnotesize Zero-neighbour isolates excluded.} \\
\bottomrule
\end{tabular}

  \end{table}

\subsection{Do missing public sector workers explain why LM breaks down at smaller geographies?}
\label{sec:decomp}

Recall that unions representing only state, county, and municipal workers are not required to file LM forms with OLMS.  LM data therefore miss some public sector workers. The extent of this gap is difficult to pin down because many unions represent a mix of both public and private sector workers but report the combined total in LM forms. One plausible reason why the LM and MrP estimates diverge at the substate level could be differential public sector employment (and unionization) across different geographies, differences that average out at the state level.  If this is the case then the LM data may well be a useful estimate of substate \emph{private sector} unionization, even as it misses government workers.

To test this conjecture, we look at two quantities.  First is the distribution of the MrP-LM union density \emph{gap}.  If missing public sector workers in the LM data account for this gap, there should be virtually no negative values in this distribution.  Second, we regress the MrP-LM density gap on the state and local government employment share.  If the MrP-LM gap results from missing state and local government union members then the gap will be bigger in geographies with more state and local government workers.  We focus on counties and states, as these are political units that have their own employees, hiring practices, and unions (unlike CDs and CZs).  

\begin{figure}[htbp]
  \centering
  \includegraphics[width=\linewidth]{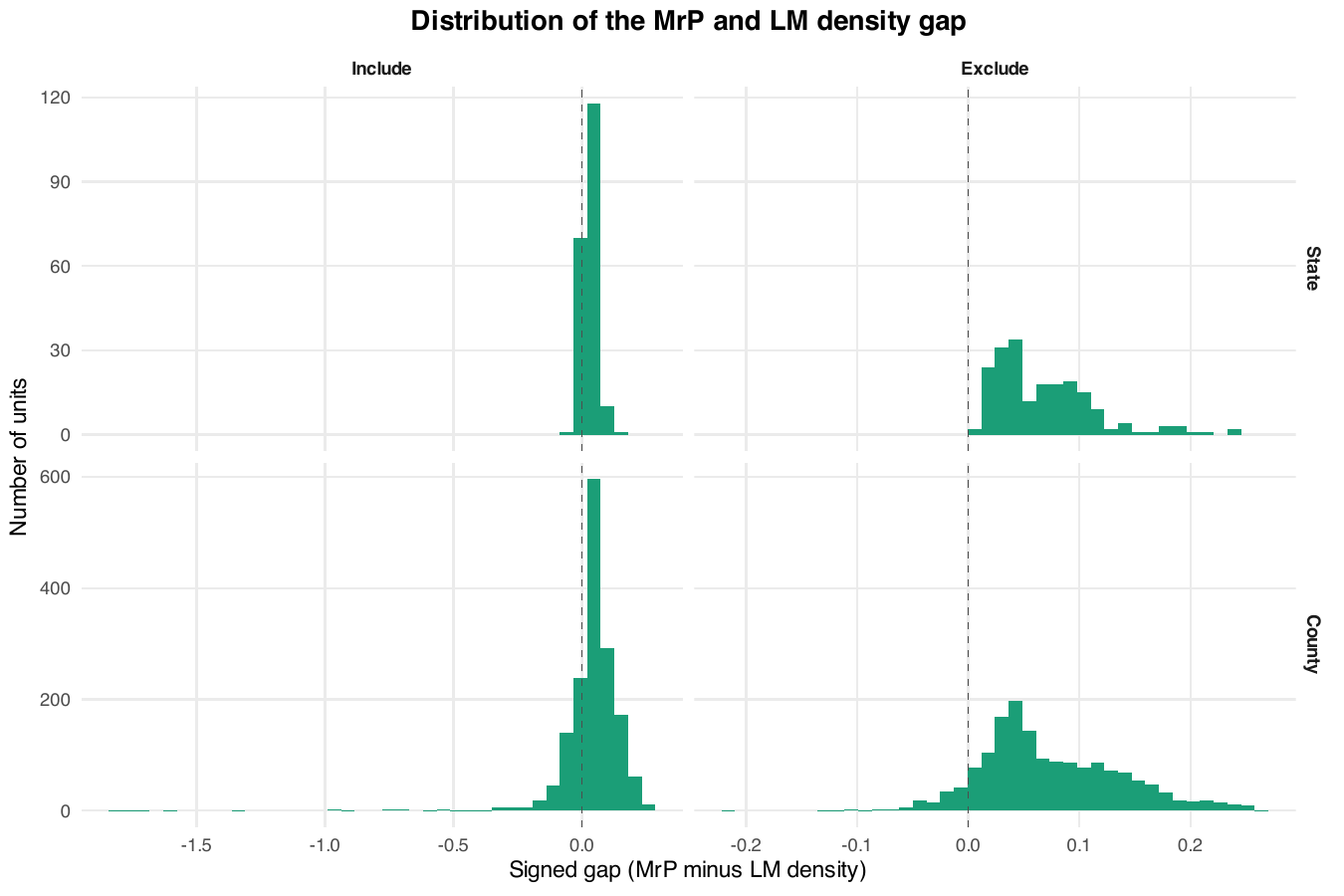}
  \caption{Distribution of the signed MrP--LM union density gap (MrP density minus LM density) at the state and county levels, for the inclusive and exclusive LM membership definitions, pooled over 2000, 2007, 2014, and 2019.}
  \label{fig:gap_signed_distribution}
\end{figure}

Figure~\ref{fig:gap_signed_distribution} displays the distributions of the MrP-LM density gap at the state and county levels by LM definition and pooling all years.  At the state level, the inclusive LM estimates are tightly clustered around the MrP estimates, but there are 14 of 200 state-years in which the gap is negative.  When using the exclusive version of the LM data MrP estimates are greater than LM estimates for all state-years, but there is a wider distribution of gaps.  Looking at the county level, however, we see a different pattern: the gap with the LM-inclusive density displays a long negative tail, reflecting the lumpy estimates just discussed.  The exclusive version of the LM data mitigates this problem substantially, but there are still six county-years in which the LM density estimates exceed those from MrP by over 10 percentage points.  This is not consistent with a purely missing public sector explanation.   

Turning to a regression approach, we source state and local government employment shares from the relevant ACS.  We pool data across all four years and estimate models with fixed effects by year, clustering standard errors on the geography in question.  We also add two covariates: urbanization (urban population share at the state level, NCHS 6-level urban-rural codes for counties) and log employment to account for the possible concentration of public sector employees in bigger and more densely populated areas.  Finally, as states vary dramatically in their treatment of public-sector unions, we also include a model that interacts this share with an indicator for the states in which public-sector collective bargaining is effectively prohibited.\footnote{We used the coding reported in \citet{brannickWorkerFreedomStates2019}. Georgia, Indiana, North Carolina, South Carolina, Tennessee, Texas, and Virginia were given the highest grade in the restriction of public sector unionization.} We treat this indicator as constant over our window, so it is identified only in the county-level models.  We estimate these models separately for the inclusive and exclusive versions of the LM data.

\begin{table}[htbp]
\centering
\caption{Regression of the signed MrP--LM density gap (in percentage points) on the state and local government employment share (in percentage points). Standard errors clustered on the geographic unit in parentheses. Estimated separately at the state and county levels for the inclusive and exclusive LM definitions. *** $p<0.001$, ** $p<0.01$, * $p<0.05$, . $p<0.1$. CB = collective bargaining.}
\label{tab:gap_reg_signed}
\small\setlength{\tabcolsep}{5pt}
\begin{tabular}{l rr rr rr}
\toprule
 & \multicolumn{2}{c}{State} & \multicolumn{2}{c}{County} & \multicolumn{2}{c}{County $\times$ CB ban} \\
\cmidrule(lr){2-3}\cmidrule(lr){4-5}\cmidrule(lr){6-7}
 & Include & Exclude & Include & Exclude & Include & Exclude \\
\midrule
State/local gov.\ \% & -0.033 & 0.298 & 0.164 & 0.547*** & 0.146 & 0.581*** \\
 & (0.230) & (0.431) & (0.381) & (0.077) & (0.465) & (0.082) \\
$\times$ CB ban &  &  &  &  & -0.048 & -0.405*** \\
 &  &  &  &  & (0.454) & (0.099) \\
Public sector CB banned &  &  &  &  & -3.159 & -0.990 \\
 &  &  &  &  & (5.627) & (1.363) \\
Urbanicity & 0.056* & 0.162** & 0.673 & 0.351 & 0.406 & -0.063 \\
 & (0.022) & (0.056) & (0.525) & (0.350) & (0.530) & (0.311) \\
Log employment & -1.670*** & -1.026 & -2.357** & 0.814* & -2.863*** & -0.004 \\
 & (0.299) & (0.872) & (0.878) & (0.413) & (0.851) & (0.368) \\
\addlinespace
Year fixed effects & Yes & Yes & Yes & Yes & Yes & Yes \\
\addlinespace
Observations & 200 & 200 & 1,613 & 1,613 & 1,613 & 1,613 \\
Clusters (units) & 50 & 50 & 475 & 475 & 475 & 475 \\
W/in $R^2$ & 0.307 & 0.205 & 0.040 & 0.130 & 0.053 & 0.318 \\
\bottomrule
\end{tabular}

\end{table}

Table~\ref{tab:gap_reg_signed} displays results.  At the state level, there is no evidence that the size of the public sector accounts for the MrP-LM gap.  But looking at the county level, we do see some evidence that more local/state public sector workers are associated with a larger gap, but only when using the exclusive LM data.  A 1pp higher level of public sector employment in a county is associated with about 0.55pp larger MrP-LM gap. The third set of models shows that this relationship all but disappears in states where public sector unions are rare, consistent with the notion that part of the gap is due to missing public sector union members.  On a model $R^2$ basis, state/local public sector employment explains about 12\% of the MrP-LM gap when using the exclusive LM data (Appendix Table~\ref{tab:gap_reg_bare_signed}).\footnote{Appendix~\ref{app:gap-regressions} reports the absolute-gap analog (Table~\ref{tab:gap_reg_absolute}) and county-level scatter diagnostics (Figure~\ref{fig:gap_pubsec_scatter}).}   

\subsection{Summary}

The exploration here suggests that the correspondence between LM-derived union density and the MrP benchmark breaks down dramatically at smaller geographies.  Some of this can be explained by lumpy reporting, in which the location of union offices (and hence member reporting) is concentrated in ways that do not coincide with workers' residential patterns as reflected in the CPS and ACS.  Lumpy reporting gives rise to impossible density values at the CD and county levels when not addressed.  But even after doing so, the MrP-LM correspondence improves only slightly.  An additional source of the divergence is that some unions include unemployed and retired members on their LM forms while others may not.  This variation, inconsistently reported on the LM forms in the case of retirees, is less likely to average out at the sub-state level. There is some evidence that MrP-LM divergence is partly due to missing public sector union members, but this explanation is limited to the exclusive version of the LM data and only accounts for about 12\% of the gap.   Overall, LM-generated estimates of union density below the state level show only a weak correlation with a reasonable benchmark of the quantity they purport to measure.     

\section{Do the differences matter?}
\label{sec:consequences}

LM-based estimates of union density are only weakly correlated with the MrP benchmark at the CD, county, and 1990 CZ levels.  But does this matter for applied work?  We provide one simple example demonstrating that it does. Using MrP and the different versions of the LM data produce quite different results that would lead to different inferences under standard practice. The goal here is to demonstrate measure sensitivity, holding the design fixed.  We are not attempting to make a substantive point.

We turn to a simplified version of the well-known ``China shock'' regression, first popularized in \citet{autorChinaSyndromeLocal2013}.  Specifically, we estimate the reduced form model that regresses changes in commuting-zone level union density on a shift-share measure of Chinese import exposure.  The model can be expressed as follows:

\begin{eqnarray}
  \Delta D^{m}_{c,W}\;&\equiv&\;100\left(D^{m}_{c,e_{W}}-D^{m}_{c,s_{W}}\right),\\
  \Delta D^{m}_{c,W}
    \;&=&\;
    \alpha_{W}
    + \beta_{W}\,\Delta\mathrm{IPW}_{c,W}
    + \gamma_{W}\,D^{m}_{c,2000}
    + \mathbf{x}_{c}^{\prime}\boldsymbol{\delta}_{W}
    + \varepsilon_{c,W}
    \label{eq:china_shock}
  \end{eqnarray}

\noindent where $c$ indexes 1990 commuting zones (the 48 contiguous states; the Washington, DC commuting zone is included and assigned to Maryland) and $W$ is the differencing window $[s_W,e_W]$. The dependent variable $\Delta D^{m}_{c,W}$ is the change in union density (measure $m$) over the window, in percentage points, where we consider the MrP benchmark and the inclusive and exclusive LM versions.  $\Delta\mathrm{IPW}_{c,W}$ is the ADH shift-share measure of the CZ $c$'s exposure to Chinese imports per worker in window $W$.\footnote{We use the change in Chinese manufacturing exports to other comparable economies allocated to US CZs based on the relative size of the CZ manufacturing labor force by industry.} $D^{m}_{c,2000}$ is the zone's year-2000 density in the same measure, a baseline level of unionization.\footnote{\citet{pierceSurprisinglySwiftDecline2016} and \citet{ahlquistEffectsImportCompetition2023a} show that the size of the China shock at the industry level is endogenous to pre-shock unionization levels.} There may also be additional covariates $\mathbf{x}_{c}$, although we relegate those specifications to the Appendix. Each regression is estimated by weighted least squares with weights equal to the commuting zone's year-2000 population and standard errors clustered by state, following ADH.  We report the two outcome windows, 2000--2007 and 2007--2014, and fit the specification separately for each density measure.

\begin{table}[htbp]
\centering
\caption{China import exposure and the change in commuting-zone union density, by density measure. Weighted OLS with standard errors clustered by state in parentheses. The dependent variable is the change in union density over the window in percentage points. ** $p<0.01$, * $p<0.05$.}
\label{tab:china_shock_main}
\begin{tabular}{l ccc}
\toprule
 & MrP & LM include & LM exclude \\
\midrule
\multicolumn{4}{l}{\textit{2000--2007}} \\
\addlinespace
$\Delta\mathrm{IPW}_{o}$ & 6.650** & 10.458* & 1.962* \\
 & (2.108) & (3.920) & (0.953) \\
Union density, 2000 & -0.002 & -0.011 & -0.223** \\
 & (0.028) & (0.026) & (0.028) \\
\addlinespace
Observations & 669 & 669 & 669 \\
Clusters & 48 & 48 & 48 \\
R-squared & 0.100 & 0.081 & 0.383 \\
\addlinespace
\multicolumn{4}{l}{\textit{2007--2014}} \\
\addlinespace
$\Delta\mathrm{IPW}_{o}$ & 1.761 & -0.090 & -0.417 \\
 & (0.935) & (0.906) & (0.375) \\
Union density, 2000 & -0.084* & -0.032 & -0.136** \\
 & (0.041) & (0.043) & (0.019) \\
\addlinespace
Observations & 669 & 669 & 669 \\
Clusters & 48 & 48 & 48 \\
R-squared & 0.192 & 0.020 & 0.195 \\
\midrule
\multicolumn{4}{@{}l@{}}{\footnotesize Standard errors clustered by state.} \\
\multicolumn{4}{@{}l@{}}{\footnotesize ** p < 0.01, * p < 0.05} \\
\multicolumn{4}{@{}l@{}}{\footnotesize 669 zones in the 48 contiguous states} \\
\bottomrule
\end{tabular}

\end{table}

Table~\ref{tab:china_shock_main} reports results.\footnote{The specifications with covariates (college share, skill share, capital--labor ratio, right-to-work status) and version using the NTR-gap exposure measure \citep{pierceSurprisinglySwiftDecline2016}, appear in Appendix~\ref{app:china-shock}. They produce similar conclusions.}   In the 2000--07 window, the different estimates of CZ union density produce quite different results.  Using the MrP benchmark shows that a one standard deviation increase in Chinese import exposure increases expected union density by about 0.48 percentage points.\footnote{As unionization is declining in this period, it is better said that the increase in exposure reduced the expected decline.}  The models using LM-derived density differ dramatically.  Using the inclusive definition, the estimated effect is about 1.6 times as large as the MrP benchmark, implying an expected density change of +0.75 percentage points for a standard deviation increase in import exposure.  Using the LM-exclude version, we see an effect of the China shock that is about 30\% as large as for the MrP benchmark.  All these estimates would be declared significant under standard practice.

Estimates from the 2007--14 period diverge even further.  The MrP-based model shows a China shock effect smaller than that estimated in 2000--07 but still positive with $p = 0.06$.  But in both the LM-based models, the coefficients for Chinese import exposure have switched signs and are now indistinguishable from zero. Finally, across both estimation windows, note that the estimated relationship between 2000 density and the change as well as model fit statistics differ dramatically, depending on which source of union density data is used.  In short, the MrP--LM divergence below the state level is analytically and substantively consequential.  Given the well-understood properties of the CPS and its close correspondence with MrP, the substate divergence between MrP and LM density estimates suggests that the LM data are not reliable in the ways we may have thought.

\section{Implications for research practice}
\label{sec:implications}

Our findings have practical implications for researchers trying to construct measures of substate unionization from data reported on LM forms. LM data are a remarkable resource but the data are messy and geolocated membership may indicate something quite different from what a researcher wants to measure.  The gap between survey reports and union-reported membership grows with the geographic resolution. Five recommendations follow.

First, do not use aggregated LM-reported membership to estimate union membership below the state level without strong scope conditions. At the state level, LM and our CPS-anchored benchmark agree well. Below the state level they do not: the correlation with the benchmark falls by more than half and never rises above roughly 0.49 at the CD, CZ, or county level, in any year we examine. A researcher who needs county, commuting-zone, or district density is, on our evidence, working with a measure only weakly related to the quantity of interest. Using LM data to estimate regional quantities could still be useful if concentrating on a relatively small number of geographic units or affiliates of specific unions.  These concerns are especially pronounced when studying social or political outcomes in which workers' geography of \emph{residence} is of prime importance. A union's office location is arguably a better geographic proxy for the location of work than for neighborhoods and voting. 

Second, our findings suggest caution in geolocating and aggregating other LM-reported quantities (expenditures, assets) to sub-state levels.  We do not have data that permit a similar validation exercise for financial variables, but the lumpiness in membership does give pause.  On the other hand, financial quantities are subject to auditing requirements and investigation by OLMS and the Department of Justice in ways that membership reports are not \citep{downeyCongressionalElectionsUnion2023}.  Our caution is specifically about attributing members or other resources to small geographies, asking the filing address to stand in for where resources are deployed.   

Third, inclusion, filtering and cleaning decisions appear to be particularly consequential when working with LM data.  Using the OLMS definition of local union organizations is but one example.  Decisions about extreme and missing values and unions sharing addresses matter, but simple exclusion is not necessarily the correct decision in all applications. But our work with the inclusive vs. exclusive LM data does suggest that simply smoothing out LM-reported members in space is not a good solution. Any work using LM-reported quantities should report how conclusions are sensitive to different data filtering and cleaning decisions. Our China shock exercise shows this is not a hypothetical worry: the same regression, run on the same design, yields import-exposure coefficients that change magnitude, cross conventional significance thresholds, and switch sign depending only on which density measure is used.

Fourth, multilevel regression with poststratification appears to perform adequately for generating model-based substate benchmarks for our purposes here. Where the ACS or Census can populate poststratification tables, MrP is one instrument for substate density estimates.  But this tool also comes with real limitations and concerns.  First, MrP inherits the known limitations and problems of the CPS, so those issues remain.  Second, MrP, especially when using a shrinkage estimator as we did here, will, by design, somewhat understates within-state variation in union membership  relative to what we can see in the CPS.  Third, MrP can be computationally expensive and the data needed for post-stratification are only intermittently available.  We lean on detailed ACS five-year tables which only become available in 2007. Extending this approach to the 1990s or filling the gaps between the decennial census and the ACS era requires additional modeling assumptions we have not investigated. Generating a complete MrP-based time series will require more extensive dynamic modeling \citep{gelmanUsingMultilevelRegression}.  Smaller geographies, such as counties, will be even more difficult. 

Fifth, use LM data for the questions they are best-suited to answer: those about the union \emph{organizations} themselves. LM forms are an organization-level census: they record the birth, death, location, affiliation, leadership, and finances of local unions with a coverage and consistency no survey matches.  \citet{wilmersLaborUnionsActivist2017, holmesDynamicsUnionOrganization2007, mastersFinancialStatusNational2022} are all examples here. \citet{becherReducingUnequalRepresentation2021} are careful to point out that they use members as a proxy for organizational strength (as opposed to geolocated voters). Examples for future research include looking at how various economic shocks affect the birth and death of union organizations or the turnover and career paths of union officers.

\section{Conclusion}
\label{sec:conclusion}

Researchers interested in US unions have been turning to administrative data, specifically annual forms many unions file with the US Department of Labor.  Data from these forms have been used to study a variety of outcomes, but one common application is in estimating substate union membership levels.  Researchers filter the administrative forms to avoid double counting, geolocate union offices, and then aggregate union-reported membership to the county, commuting zone, and Congressional district.  We attempted to put this practice on sounder footing by validating these substate estimates using a triangulation strategy based on multilevel regression with poststratification.

Although we showed that the MrP estimates produced an adequate benchmark, we also found that the correlation between the LM-based estimates and this benchmark collapsed below the state level.  Moreover, this did not change across different types and scales of substate geographies or over time.  Membership data as reported on LM forms displayed considerable lumpiness that distorted estimates in smaller areas of aggregation.  Choices about how to filter and clean these data were also consequential, but failed to restore a strong correlation with the benchmark.  We found some evidence that missing public sector unions in the administrative data explained about 12\% of the gap between the MrP benchmarks and the LM-based estimates, substantial but by no means overwhelming.    

Finally, we showed that this distinction matters for applied research.  Estimating the standard China shock shift-share regression on CZ-level union density for the 2000--2014 period, we observed results that differ dramatically depending on which dataset we used to calculate substate density.  Overall, we conclude that applied researchers can rely on LM data to study organization-level questions, but aggregating LM-reported membership is \emph{not} a reliable method for estimating sub-state union density.  Used for what they are, LM forms remain among the richest sources available on American labor organizations; used to map members onto small geographies, they can mislead in ways that matter for the conclusions researchers draw.

 \bibliographystyle{apsr}
 \bibliography{LM_bib}

\appendix

\section{Model selection: candidates, tuning, and performance}
\label{app:model-selection}

\subsection{Candidate models}
\label{app:candidates}

We considered seven candidate models, all fit to the same CPS training split for
each comparison year (2007 and 2019) and scored on the same held-out test split.

\begin{enumerate}
  \item \textbf{Null model.} Logistic model with a state random intercept only,
        $\text{union} \sim (1 \mid \text{state})$, fit with \texttt{stan\_glmer}.
  \item \textbf{Naive Bayes.} Kernel naive Bayes classifier over the individual
        predictors.
  \item \textbf{Demographic model.} Multilevel logit with demographic and industry-sector
        main effects and a state random intercept.
  \item \textbf{Hierarchical logit.} The demographic model plus a model for state effects using right-to-work
        exposure and Census region.
  \item \textbf{Saturated hierarchical logit.} The hierarchical specification plus a
        random effect for the fully crossed demographic cell
        (age $\times$ race $\times$ sex $\times$ education). Industry sector enters as a
        main effect only, and is deliberately \emph{not} folded into the
        four-way interaction, to hold down the interaction cell count and the
        associated convergence risk.
  \item \textbf{Random forest.} Probability forest (500 trees) over the same
        predictors.
  \item \textbf{XGBoost.} Gradient-boosted trees over the same predictors.
\end{enumerate}

The individual predictors are age, race, sex, education, and industry-sector.  This last variable was constructed by assigning each worker to one of five mutually exclusive industry-sector categories by a strict precedence rule: anyone whose class of worker is federal, state, or local government is coded public; everyone else is classified from their 1990-basis industry code as construction (which also absorbs extraction and mining), manufacturing, or transportation/communications/utilities; and all remaining private-sector workers fall into a residual other category. An eighth model, a random forest that also receives the state as a predictor, is reported alongside the candidates for comparison; it is the only model given the state directly and was not a candidate for selection.

The comparison was run on 2007 and 2019. The selected hierarchical logit was then fit to the full pooled CPS sample (training and test splits combined) for each of the four analysis years, 2000, 2007, 2014, and 2019.

\subsection{Tuning}
\label{app:tuning}

\paragraph{Gradient-boosted trees.} Seven hyperparameters (number of trees, tree
depth, minimum node size, loss reduction, subsample size, \texttt{mtry}, and
learning rate) were tuned by a 15-point space-filling grid followed by 20
iterations of Bayesian optimization, over three-fold cross-validation of the training split, maximizing cross-validated AUC. The search was run separately for each comparison year.

\paragraph{Random forest.} \texttt{mtry} $\in \{2,\dots,5\}$ and
\texttt{min\_n} $\in \{5,\dots,20\}$ were tuned over a 16-point regular grid
with 200 trees, three-fold cross-validation, maximizing AUC; the selected
configuration was then re-fit with 500 trees. The same fitted forest produces the
individual-level predictions and the poststratified state and district estimates.

\paragraph{Bayesian models.} The null, demographic logit, hierarchical logit, and saturated
hierarchical logit were not cross-validation--tuned. The hierarchical and saturated hierarchical logits use weakly informative priors described in the main text
(coefficients $\mathcal{N}(0,1)$, intercept $\mathcal{N}(0,2)$, both autoscaled;
covariance \texttt{decov}$(0.5)$); the null and demographic models use the \texttt{rstanarm} default priors. All four were fit with four MCMC chains, 2000
iterations (1000 warmup), \texttt{adapt\_delta}\,$=0.8$. Convergence of the
final hierarchical logit fit is monitored by maximum $\hat{R}$ and minimum
effective sample size. Across the four analysis years the maximum $\hat{R}$ is at most
1.004 and the minimum effective sample size is at least 876
(Table~\ref{tab:mrp-convergence}), against the pipeline's warning
thresholds of $\hat{R}>1.05$ and an effective sample size below 400.

\begin{table}[htbp]
\centering
\caption{Convergence diagnostics for the final hierarchical logit fit, by analysis year: training sample size, maximum potential scale reduction factor ($\hat{R}$), and minimum effective sample size, over all model parameters and the log posterior.}
\label{tab:mrp-convergence}
\begin{tabular}{rrrr}
\toprule
Year & Training N & Max.\ $\hat{R}$ & Min.\ ESS \\
\midrule
2000 & 497,309 & 1.004 & 1026 \\
2007 & 530,581 & 1.002 & 1078 \\
2014 & 498,234 & 1.002 & 876 \\
2019 & 448,253 & 1.002 & 1121 \\
\bottomrule
\end{tabular}

\end{table}

\subsection{Split structure}
\label{app:splits}

For each comparison year the CPS sample pools three consecutive years centered on that year (2006--2008 for 2007; 2018--2020 for 2019) and was divided 80/20 into training and test sets, stratified on union membership. Hyperparameter tuning used three-fold cross-validation \emph{within} the training split.

\subsection{Evaluation criteria and per-model performance}
\label{app:performance}

\paragraph{Individual level (held-out test split).} AUC was the primary metric, but we also looked at balanced
accuracy, F1, precision, and recall. In both comparison years the AUC of the strongest models falls within roughly 0.80--0.83, and the hierarchical logit sits within that group. Full values are in
Table~\ref{tab:individual-metrics}.

\begin{table}[htbp]
\centering
\caption{Individual-level held-out performance by model, 2007 and 2019: area under the ROC curve (AUC), balanced accuracy, F1, precision, and recall. The last four classify at the training-sample union membership rate (0.122 in 2007, 0.103 in 2019).}
\label{tab:individual-metrics}
\begin{tabular}{lrrrrr}
\toprule
Model & AUC & Balanced accuracy & F1 & Precision & Recall \\
\midrule
\multicolumn{6}{l}{\textit{2007 (CPS 2006--2008; train 424,464 / test 106,117)}} \\
Null model & 0.655 & 0.614 & 0.273 & 0.168 & 0.728 \\
Naive Bayes & 0.799 & 0.723 & 0.405 & 0.289 & 0.676 \\
Demographic model & 0.815 & 0.737 & 0.415 & 0.292 & 0.713 \\
Hierarchical logit (selected) & 0.815 & 0.737 & 0.415 & 0.292 & 0.713 \\
Saturated hier. logit & 0.820 & 0.742 & 0.418 & 0.293 & 0.726 \\
Random forest & 0.826 & 0.751 & 0.433 & 0.308 & 0.729 \\
XGBoost & 0.828 & 0.753 & 0.434 & 0.309 & 0.733 \\
\multicolumn{6}{l}{\textit{Comparison only, not a selection candidate}} \\
Random forest + state ID & 0.831 & 0.754 & 0.433 & 0.306 & 0.741 \\
\addlinespace
\multicolumn{6}{l}{\textit{2019 (CPS 2018--2020; train 358,601 / test 89,652)}} \\
Null model & 0.663 & 0.624 & 0.250 & 0.152 & 0.690 \\
Naive Bayes & 0.765 & 0.662 & 0.267 & 0.159 & 0.833 \\
Demographic model & 0.803 & 0.726 & 0.369 & 0.252 & 0.685 \\
Hierarchical logit (selected) & 0.803 & 0.726 & 0.369 & 0.252 & 0.685 \\
Saturated hier. logit & 0.808 & 0.730 & 0.369 & 0.250 & 0.701 \\
Random forest & 0.815 & 0.737 & 0.373 & 0.251 & 0.721 \\
XGBoost & 0.817 & 0.741 & 0.385 & 0.265 & 0.709 \\
\multicolumn{6}{l}{\textit{Comparison only, not a selection candidate}} \\
Random forest + state ID & 0.820 & 0.743 & 0.382 & 0.259 & 0.726 \\
\bottomrule
\end{tabular}

\end{table}

\paragraph{State level (poststratified estimate vs.\ CPS (Unionstats)).} Root-mean-squared error and Pearson correlation of each model's poststratified state density against the Unionstats benchmark for all workers and for the private sector (50 states and DC). Full values are in Table~\ref{tab:state-metrics}.

\begin{table}[htbp]
\centering
\caption{State-level model performance, 2007 and 2019: RMSE and Pearson correlation of each model's poststratified state density against CPS (Unionstats), for all workers and for the private sector.}
\label{tab:state-metrics}
\begin{tabular}{lrrrr}
\toprule
 & \multicolumn{2}{c}{All workers} & \multicolumn{2}{c}{Private sector} \\
\cmidrule(lr){2-3} \cmidrule(lr){4-5}
Model & RMSE & Correlation & RMSE & Correlation \\
\midrule
\multicolumn{5}{l}{\textit{2007 (CPS 2006--2008; train 424,464 / test 106,117)}} \\
Null model & 0.005 & 0.996 & 0.052 & 0.931 \\
Naive Bayes & 0.028 & 0.884 & 0.027 & 0.762 \\
Demographic model & 0.009 & 0.990 & 0.014 & 0.928 \\
Hierarchical logit (selected) & 0.009 & 0.990 & 0.014 & 0.927 \\
Saturated hier. logit & 0.009 & 0.991 & 0.014 & 0.928 \\
Random forest & 0.019 & 0.936 & 0.015 & 0.911 \\
XGBoost & 0.019 & 0.940 & 0.014 & 0.922 \\
\multicolumn{5}{l}{\textit{Comparison only, not a selection candidate}} \\
Random forest + state ID & 0.015 & 0.963 & 0.012 & 0.945 \\
\addlinespace
\multicolumn{5}{l}{\textit{2019 (CPS 2018--2020; train 358,601 / test 89,652)}} \\
Null model & 0.006 & 0.995 & 0.049 & 0.917 \\
Naive Bayes & 0.139 & 0.823 & 0.121 & 0.656 \\
Demographic model & 0.009 & 0.991 & 0.013 & 0.916 \\
Hierarchical logit (selected) & 0.009 & 0.990 & 0.013 & 0.915 \\
Saturated hier. logit & 0.009 & 0.990 & 0.013 & 0.916 \\
Random forest & 0.020 & 0.915 & 0.015 & 0.863 \\
XGBoost & 0.017 & 0.940 & 0.013 & 0.895 \\
\multicolumn{5}{l}{\textit{Comparison only, not a selection candidate}} \\
Random forest + state ID & 0.012 & 0.973 & 0.009 & 0.954 \\
\bottomrule
\end{tabular}

\end{table}

\subsection{Selection}
\label{app:selection}

Model choice weighed three criteria: individual-level discrimination on the
held-out test split (AUC), recovery of the state-level union-density aggregate
against the external Unionstats CPS reference, and the variation in
poststratified density across congressional districts within each state.
The saturated hierarchical logit differs from the hierarchical logit only by its
random effect for the fully crossed demographic cell. Its AUC is higher by 0.005
in each year (0.820 against 0.815 in 2007; 0.808 against 0.803 in 2019); the two
agree to three decimals on state-level RMSE and within-state district variation,
and to within 0.001 on state-level correlation. We evaluated the saturated
specification, judged that the interaction did not add enough on these criteria
to justify it and its computational cost, and adopted the hierarchical logit.
Individual-level discrimination is comparable across the strongest models (AUC
$\approx 0.80$--$0.83$), and the hierarchical logit is within that group. The null
model has the lowest state-level RMSE in both years; because it has no
individual-level predictors, it assigns every congressional district in a state
the same density, so its within-state standard deviation is zero by construction.
Table~\ref{tab:model-comparison} places the three criteria side by side for all
eight models.

\begin{table}[htbp]
\centering
\caption{The three selection criteria side by side, 2007 and 2019: individual-level AUC, state-level RMSE and correlation against CPS (Unionstats), and mean within-state district variation. The random forest with a state identifier was not a selection candidate.}
\label{tab:model-comparison}
\begin{tabular}{lrrrr}
\toprule
 & Individual & \multicolumn{2}{c}{State (all workers)} & Districts \\
\cmidrule(lr){2-2} \cmidrule(lr){3-4} \cmidrule(lr){5-5}
Model & AUC & RMSE & Correlation & Mean w/in-state SD \\
\midrule
\multicolumn{5}{l}{\textit{2007 (CPS 2006--2008; train 424,464 / test 106,117)}} \\
Null model & 0.655 & 0.005 & 0.996 & 0.000 \\
Naive Bayes & 0.799 & 0.028 & 0.884 & 0.008 \\
Demographic model & 0.815 & 0.009 & 0.990 & 0.010 \\
Hierarchical logit (selected) & 0.815 & 0.009 & 0.990 & 0.010 \\
Saturated hier. logit & 0.820 & 0.009 & 0.991 & 0.010 \\
Random forest & 0.826 & 0.019 & 0.936 & 0.009 \\
XGBoost & 0.828 & 0.019 & 0.940 & 0.009 \\
\multicolumn{5}{l}{\textit{Comparison only, not a selection candidate}} \\
Random forest + state ID & 0.831 & 0.015 & 0.963 & 0.008 \\
\addlinespace
\multicolumn{5}{l}{\textit{2019 (CPS 2018--2020; train 358,601 / test 89,652)}} \\
Null model & 0.663 & 0.006 & 0.995 & 0.000 \\
Naive Bayes & 0.765 & 0.139 & 0.823 & 0.009 \\
Demographic model & 0.803 & 0.009 & 0.991 & 0.008 \\
Hierarchical logit (selected) & 0.803 & 0.009 & 0.990 & 0.008 \\
Saturated hier. logit & 0.808 & 0.009 & 0.990 & 0.008 \\
Random forest & 0.815 & 0.020 & 0.915 & 0.008 \\
XGBoost & 0.817 & 0.017 & 0.940 & 0.008 \\
\multicolumn{5}{l}{\textit{Comparison only, not a selection candidate}} \\
Random forest + state ID & 0.820 & 0.012 & 0.973 & 0.008 \\
\bottomrule
\end{tabular}

\end{table}

\subsection{Details of the selected hierarchical logit}\label{app:shl_selected}

The selected hierarchical logit can be expressed as follows:

\begin{align}
y_i
&\sim
\operatorname{Bernoulli}(p_i),
\qquad i=1,\ldots,n,
\label{eq:mrp-outcome}
\\[4pt]
\operatorname{logit}(p_i)
&=
\alpha_{j[i]}
+ \boldsymbol{\beta}_1^{\mathsf T}\mathrm{age}_i
+ \boldsymbol{\beta}_2^{\mathsf T}\mathrm{race}_i
+ \boldsymbol{\beta}_3^{\mathsf T}\mathrm{sex}_i
\nonumber\\
&\quad
+ \boldsymbol{\beta}_4^{\mathsf T}\mathrm{education}_i
+ \boldsymbol{\beta}_5^{\mathsf T}\mathrm{industry-sector}_i,
\label{eq:mrp-individual}
\\[4pt]
\alpha_j \mid \boldsymbol{\gamma},\sigma_{\mathrm{state}}
&\sim
\operatorname{Normal}\!\left(\mu_j,\ \sigma_{\mathrm{state}}^2\right),
\qquad j=1,\ldots,J,
\label{eq:mrp-state-distribution}
\\[4pt]
\mu_j
&=
\gamma_0
+ \gamma_1\mathrm{RTW}_j
+ \boldsymbol{\gamma}_2^{\mathsf T}\mathrm{region}_j,
\label{eq:mrp-state-mean}
\\[4pt]
\beta_q
&\sim
\operatorname{Normal}(0,\ \xi_q^2),
\qquad q=1,\ldots,Q_I,
\label{eq:mrp-individual-priors}
\\[4pt]
\gamma_r
&\sim
\operatorname{Normal}(0,\ \zeta_r^2),
\qquad r=1,\ldots,Q_S,
\label{eq:mrp-state-priors}
\\[4pt]
\gamma_0
&\sim
\operatorname{Normal}(0,\ \xi_0^2),
\label{eq:mrp-intercept-prior}
\\[4pt]
\sigma_{\mathrm{state}}^2
&\sim
\operatorname{Gamma}(1,\ 0.5).
\label{eq:mrp-state-var-prior}
\end{align}

Equation~\eqref{eq:mrp-outcome} is the individual-level outcome: $y_i = 1$ if respondent $i$ is a union member and $0$ otherwise, with $i = 1,\ldots,n$ indexing the CPS respondents. Equation~\eqref{eq:mrp-individual} expresses the linear individual-level predictor as a function of demographic predictors for which the joint distribution is available at lower geographic levels via the American Community Survey (ACS): age, race, sex, education, and industry/sector, with corresponding regression parameters $\beta$.

The state effects, $\alpha_{j[i]}$,  are modeled hierarchically in
Equations~\eqref{eq:mrp-state-distribution}--\eqref{eq:mrp-state-mean}: each
state's effect $\alpha_j$ is normally distributed about a mean $\mu_j$ that is a
linear function of state-level predictors: right-to-work coverage and census region---with coefficients $\gamma_0,\ldots,\gamma_2$
and residual standard deviation $\sigma_{\mathrm{state}}$.

Equations~\eqref{eq:mrp-individual-priors}--\eqref{eq:mrp-state-var-prior} give the
priors. The individual- and state-level coefficients receive weakly informative
normal priors; their scales $\xi_q$ (for the $Q_I$ individual-level
coefficients), $\zeta_r$ (for the $Q_S$ state-level coefficients), and $\xi_0$
(for the intercept) are set automatically and standardized to each predictor, so
that every coefficient receives a comparably weak prior regardless of its units.
The prior on the intercept $\gamma_0$ applies to its value after the predictors
have been centered at their means. The variance of the state effects,
$\sigma_{\mathrm{state}}^2$, receives
a $\operatorname{Gamma}(1, 0.5)$ prior (shape and scale parameterization),
following the default covariance-decomposition prior of \texttt{rstanarm}, which
for a single scalar random effect reduces to a Gamma prior on the effect's
variance.

\section{County-level CPS vs. LM}\label{app:CPS_county_v_LM}
\begin{figure}[htbp]
    \centering
    \includegraphics[width=\linewidth]{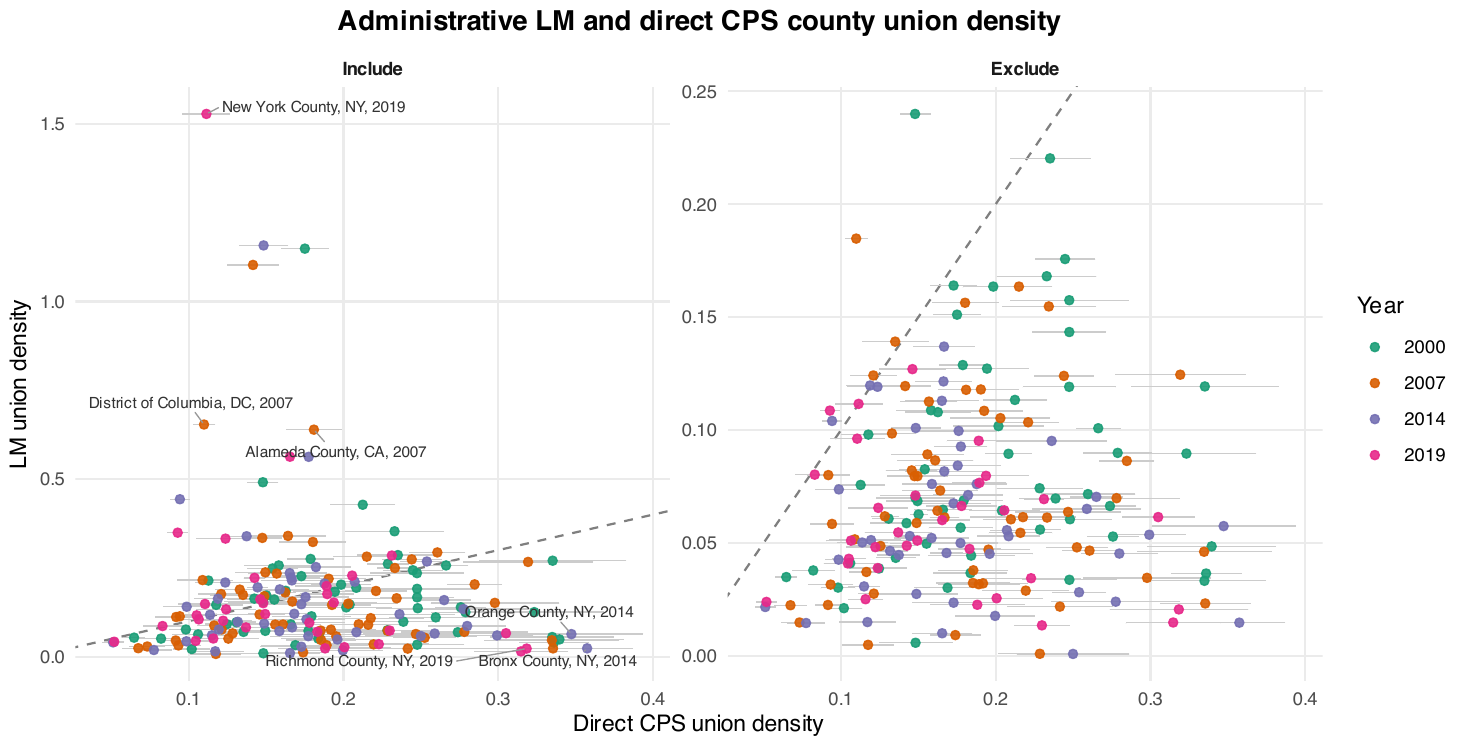}
    \caption{Administrative (LM) county union density versus direct CPS union density for 192 county-years, by membership definition. Horizontal bars are the CPS 95\% sampling interval; LM is an administrative count.}
    \label{fig:lm-vs-cps-county}
  \end{figure}

\begin{table}[htbp]
\centering
\caption{Within-state association of county MrP and LM union density with direct CPS density, by LM membership definition. Years pooled; the unit is the county-year. DC excluded.}
\label{tab:within_state_county_validation}
\begin{tabular}{l rrr}
\toprule
Estimate & Pearson & Spearman & MAE (pp) \\
\midrule
MrP & 0.577 & 0.545 & 1.44 \\
LM (include) & -0.334 & -0.146 & 10.22 \\
LM (exclude) & 0.021 & 0.049 & 2.77 \\
\midrule
\multicolumn{4}{@{}l@{}}{\footnotesize MAE = mean absolute deviation (percentage points).} \\
\multicolumn{4}{@{}l@{}}{\footnotesize 188 county-years in 81 counties, 25 states, 69 state-years.} \\
\bottomrule
\end{tabular}

\end{table}

\begin{figure}[htbp]
\centering
\includegraphics[width=\linewidth]{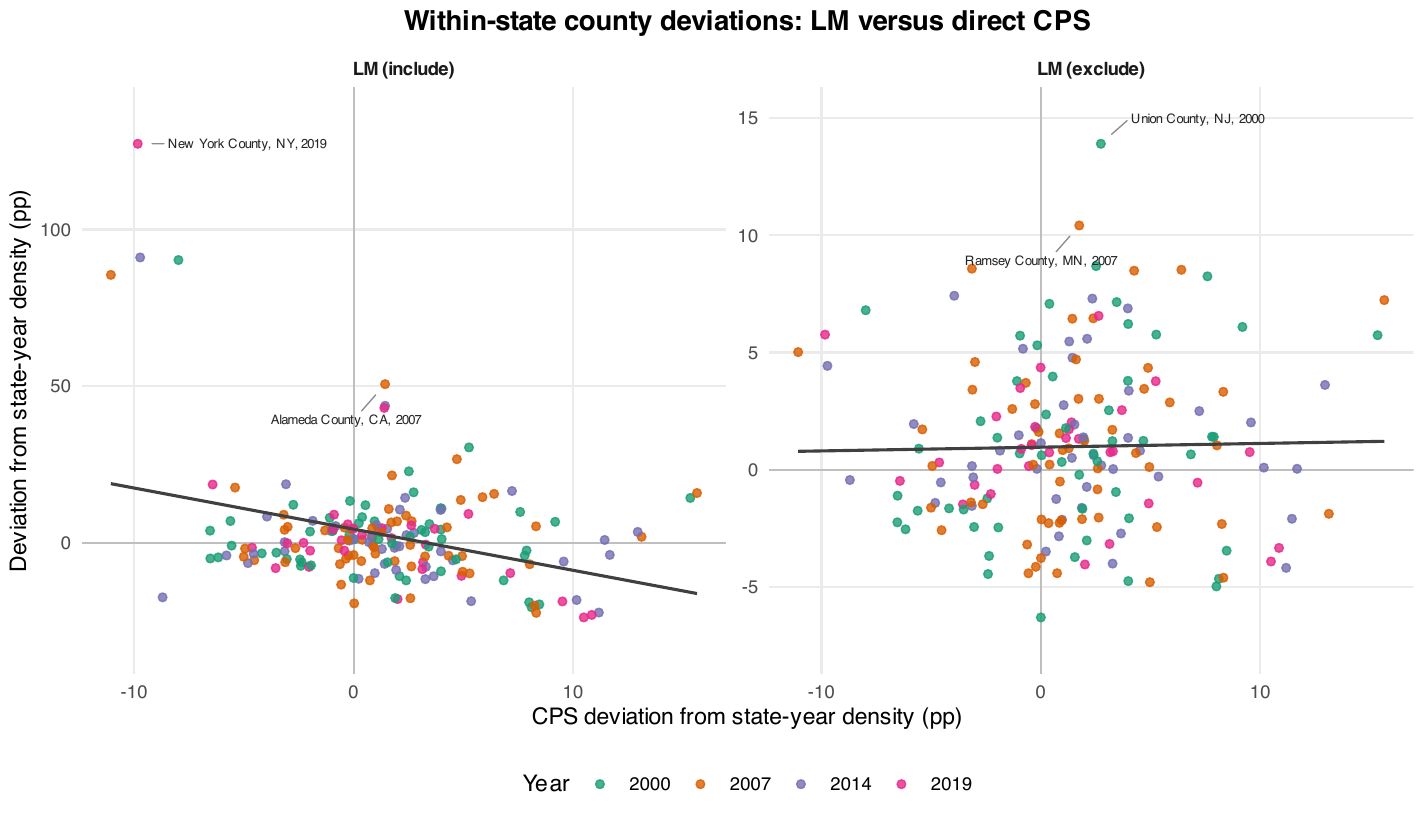}
\caption{LM within-state union density deviations versus CPS within-state deviations, by membership definition, on the same 188 county-years as Figure~\ref{fig:mrp-lm-cps-county-deviation}. Each facet carries its own fit and vertical scale.}
\label{fig:lm-cps-county-deviation}
\end{figure}

\section{Complete comparisons across years, geographies, and LM membership criteria}\label{app:complete_figs}

This appendix collects the MrP-versus-LM density comparisons underlying Section~\ref{sec:ladder}, for both the inclusive and exclusive LM membership definitions across all four years. The county-level density facets appear in the main text (Figure~\ref{fig:county_facet_mrp_lm_density}), as does the state-level 2007 inclusive comparison (Figure~\ref{fig:state_matrix_density_include_2007}).

\subsection{Substate MrP--LM density by year}

\begin{figure}[htbp]
  \centering
  \includegraphics[width=\linewidth]{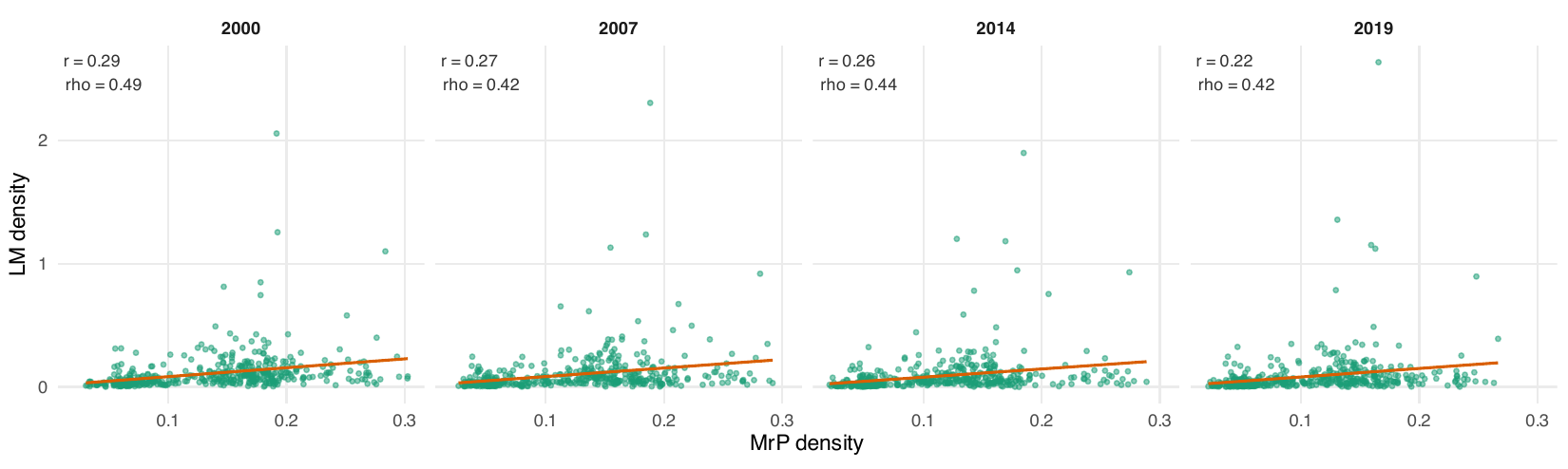}
  \caption{Union density at the congressional district level: MrP versus LM, faceted by year (2000, 2007, 2014, 2019), using the inclusive LM membership definition.}
  \label{fig:cd_facet_mrp_lm_density_include}
\end{figure}

\begin{figure}[htbp]
  \centering
  \includegraphics[width=\linewidth]{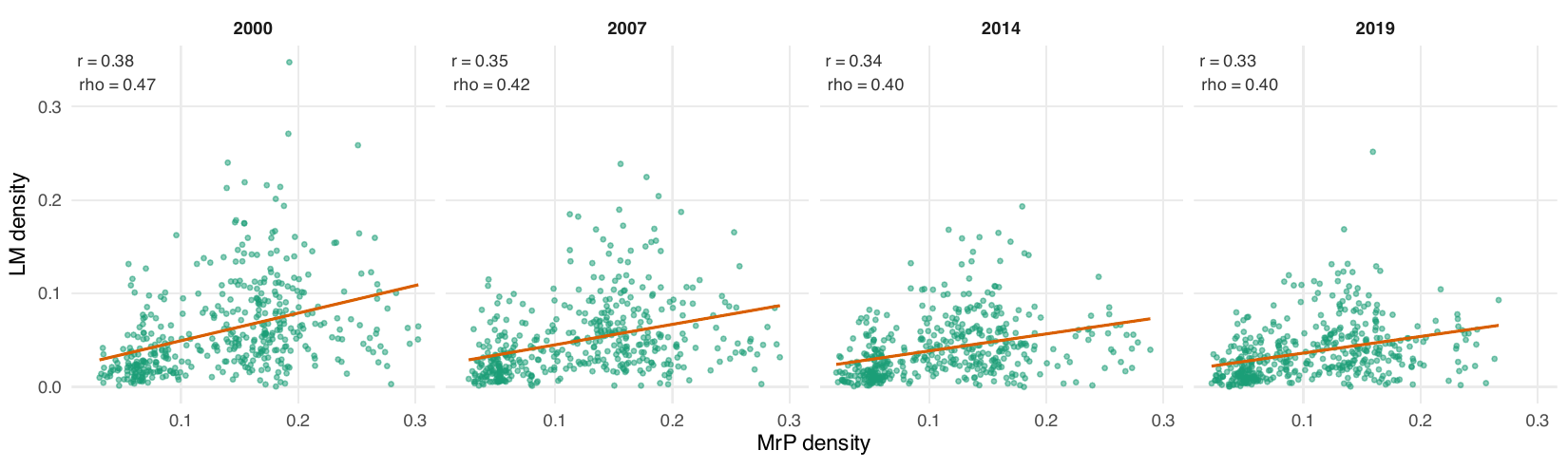}
  \caption{Union density at the congressional district level: MrP versus LM, faceted by year (2000, 2007, 2014, 2019), using the exclusive LM membership definition.}
  \label{fig:cd_facet_mrp_lm_density_exclude}
\end{figure}

\begin{figure}[htbp]
  \centering
  \includegraphics[width=\linewidth]{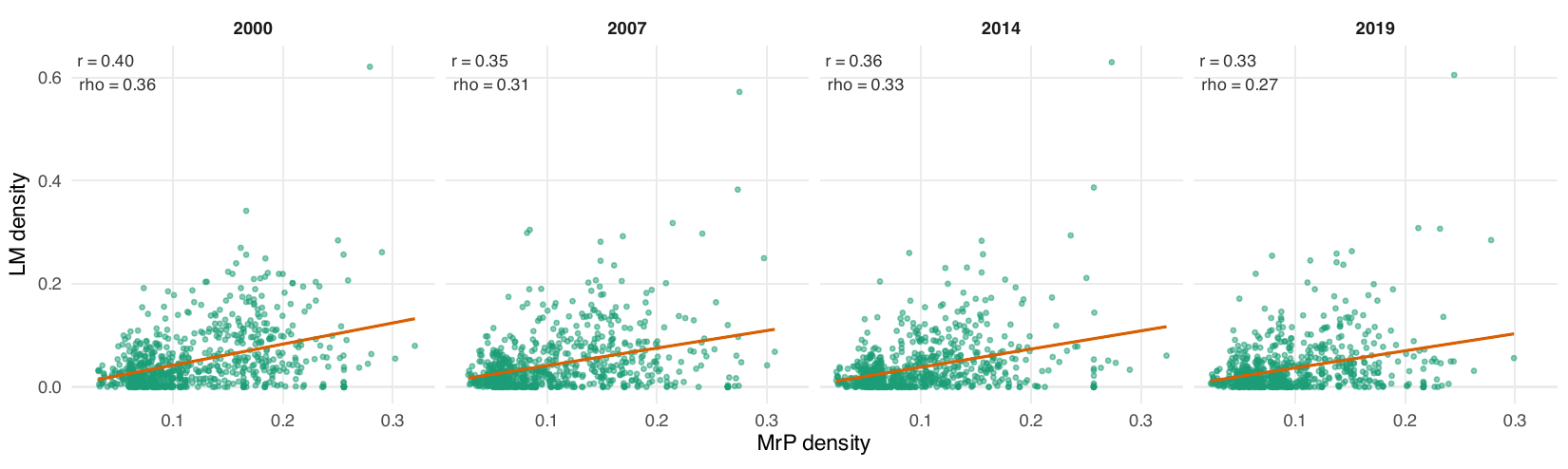}
  \caption{Union density at the commuting zone level: MrP versus LM, faceted by year (2000, 2007, 2014, 2019), using the inclusive LM membership definition.}
  \label{fig:cz_facet_mrp_lm_density_include}
\end{figure}

\begin{figure}[htbp]
  \centering
  \includegraphics[width=\linewidth]{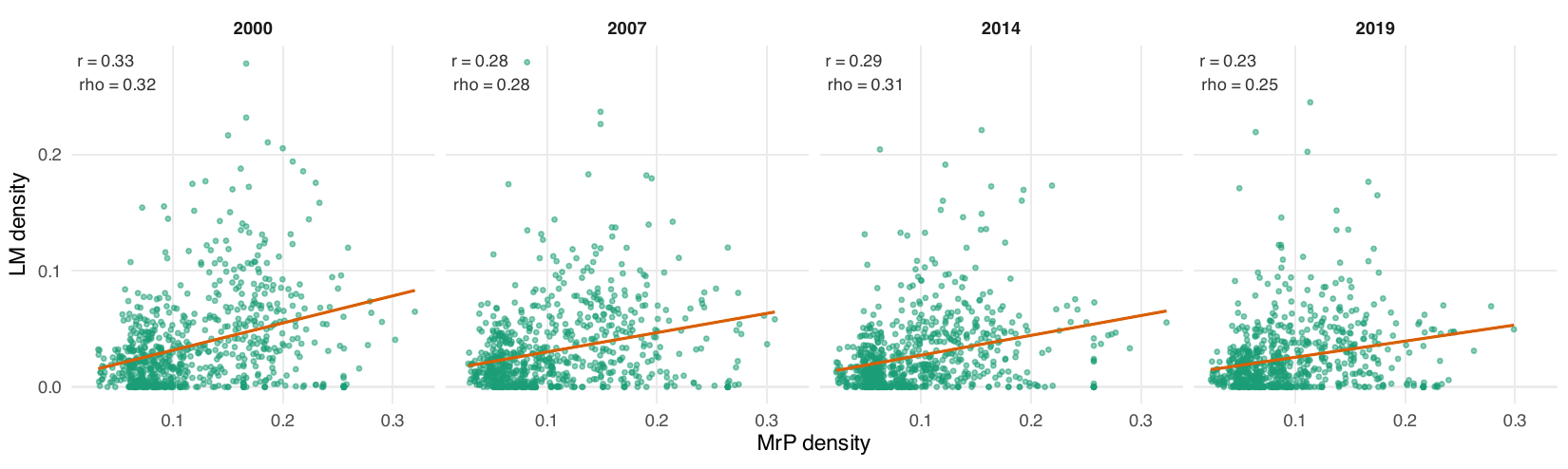}
  \caption{Union density at the commuting zone level: MrP versus LM, faceted by year (2000, 2007, 2014, 2019), using the exclusive LM membership definition.}
  \label{fig:cz_facet_mrp_lm_density_exclude}
\end{figure}

\subsection{State-level pairwise comparisons: inclusive definition}

\begin{figure}[htbp]
  \centering
  \includegraphics[width=\linewidth]{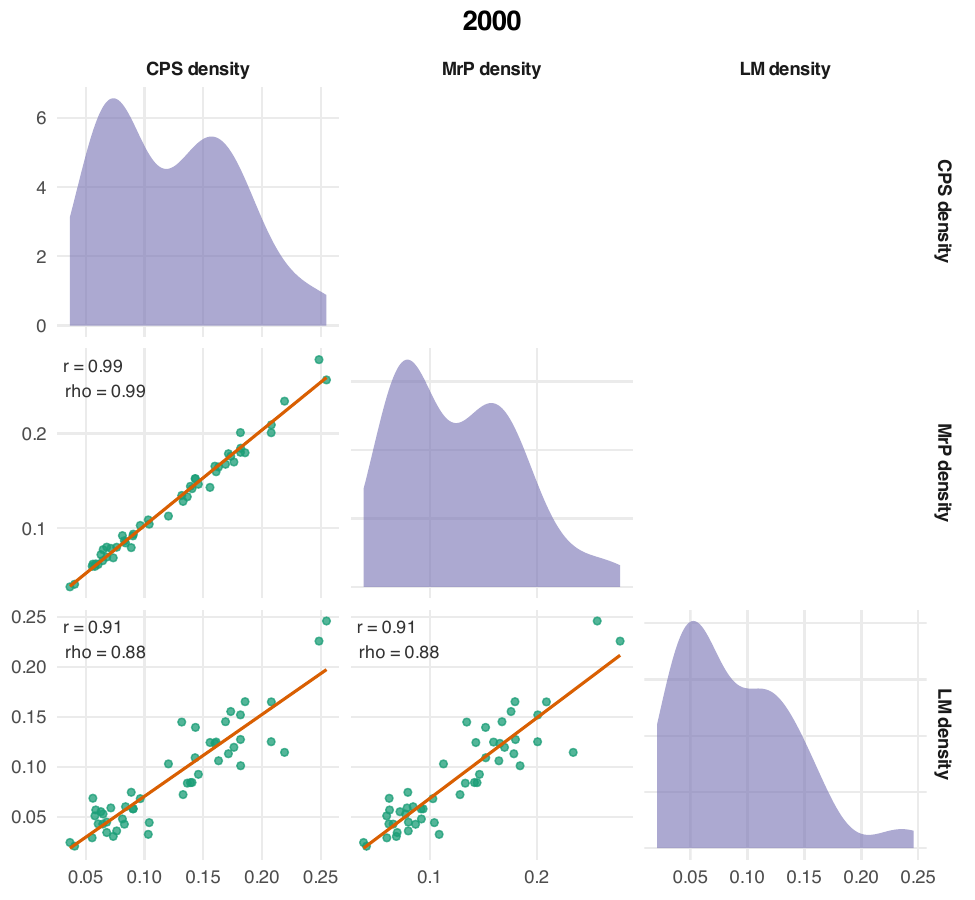}
  \caption{Union density at the state level, 2000: pairwise comparison of CPS, MrP, and LM, using the inclusive LM membership definition.}
  \label{fig:state_matrix_density_include_2000}
\end{figure}

\begin{figure}[htbp]
  \centering
  \includegraphics[width=\linewidth]{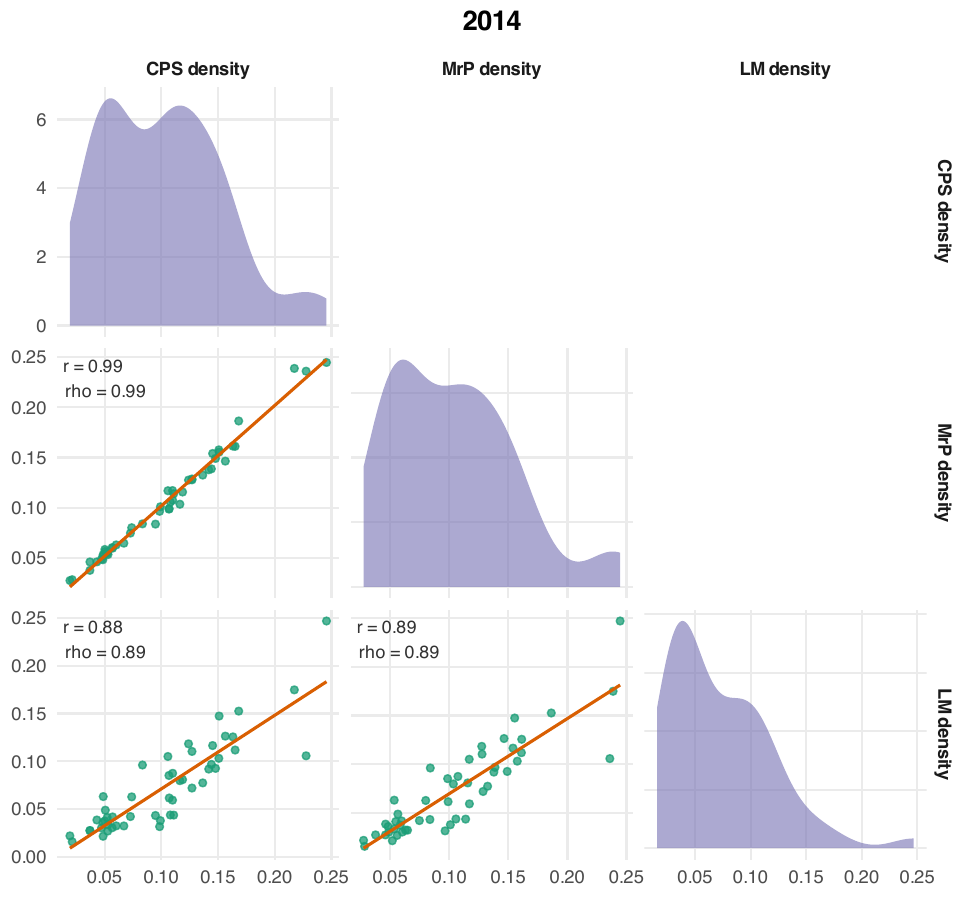}
  \caption{Union density at the state level, 2014: pairwise comparison of CPS, MrP, and LM, using the inclusive LM membership definition.}
  \label{fig:state_matrix_density_include_2014}
\end{figure}

\begin{figure}[htbp]
  \centering
  \includegraphics[width=\linewidth]{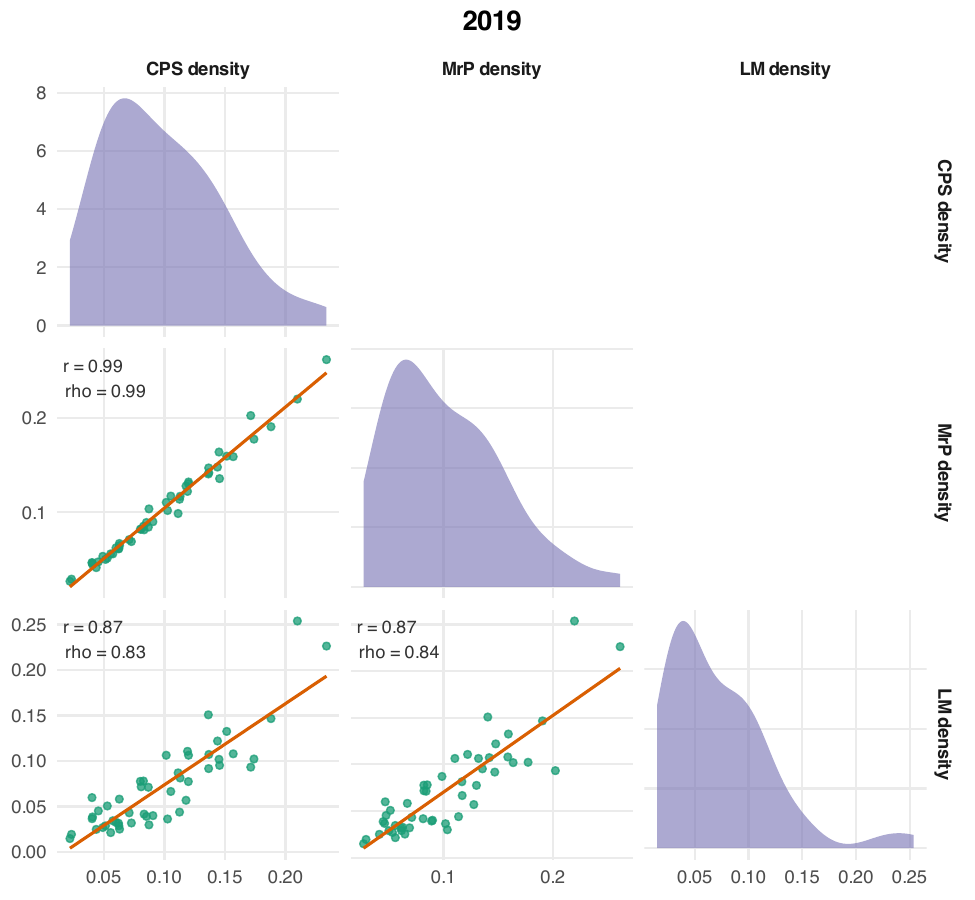}
  \caption{Union density at the state level, 2019: pairwise comparison of CPS, MrP, and LM, using the inclusive LM membership definition.}
  \label{fig:state_matrix_density_include_2019}
\end{figure}

\begin{figure}[htbp]
  \centering
  \includegraphics[width=\linewidth]{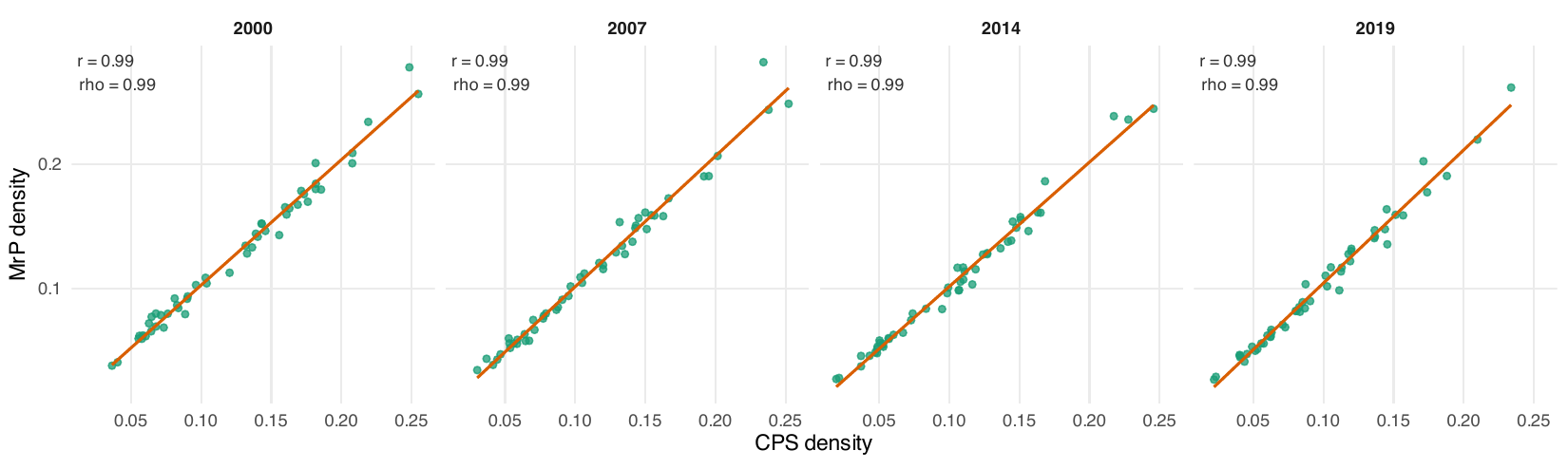}
  \caption{Union density at the state level: CPS versus MrP, faceted by year (2000, 2007, 2014, 2019), using the inclusive LM membership definition.}
  \label{fig:state_facet_density_include_cps_mrp}
\end{figure}

\begin{figure}[htbp]
  \centering
  \includegraphics[width=\linewidth]{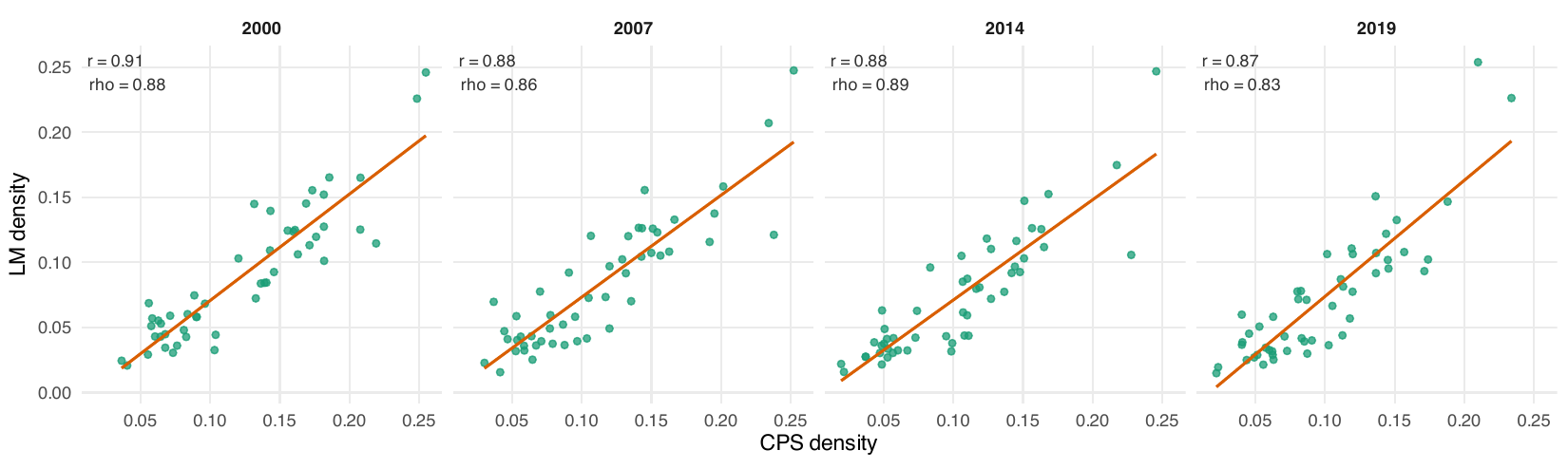}
  \caption{Union density at the state level: CPS versus LM, faceted by year (2000, 2007, 2014, 2019), using the inclusive LM membership definition.}
  \label{fig:state_facet_density_include_cps_lm}
\end{figure}

\begin{figure}[htbp]
  \centering
  \includegraphics[width=\linewidth]{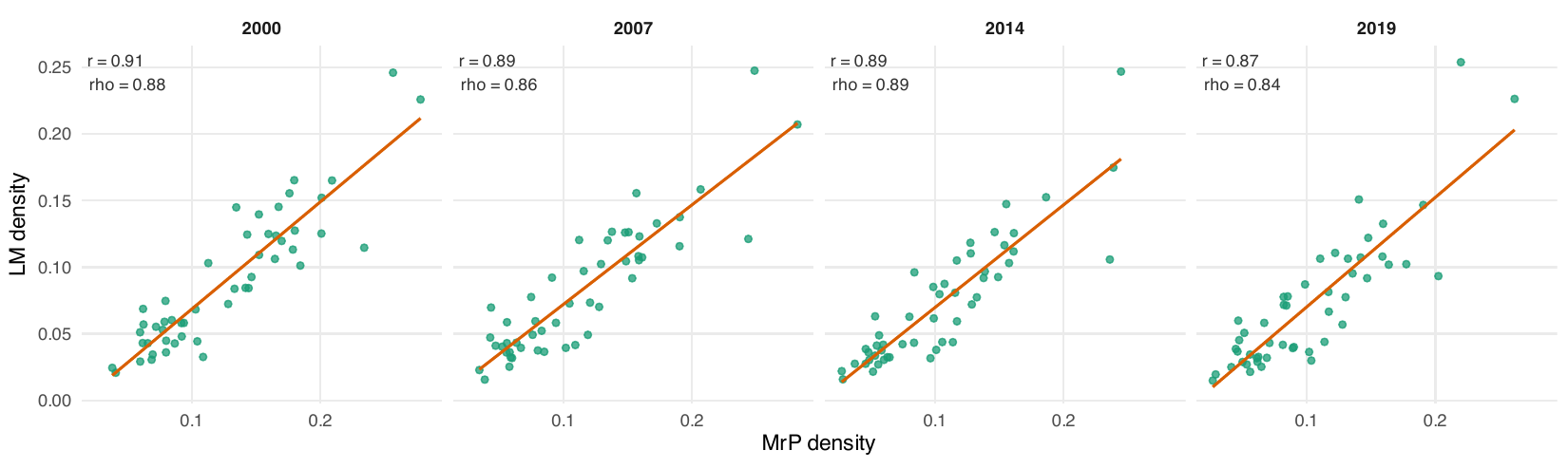}
  \caption{Union density at the state level: MrP versus LM, faceted by year (2000, 2007, 2014, 2019), using the inclusive LM membership definition.}
  \label{fig:state_facet_density_include_mrp_lm}
\end{figure}

\subsection{State-level pairwise comparisons: exclusive definition}

\begin{figure}[htbp]
  \centering
  \includegraphics[width=\linewidth]{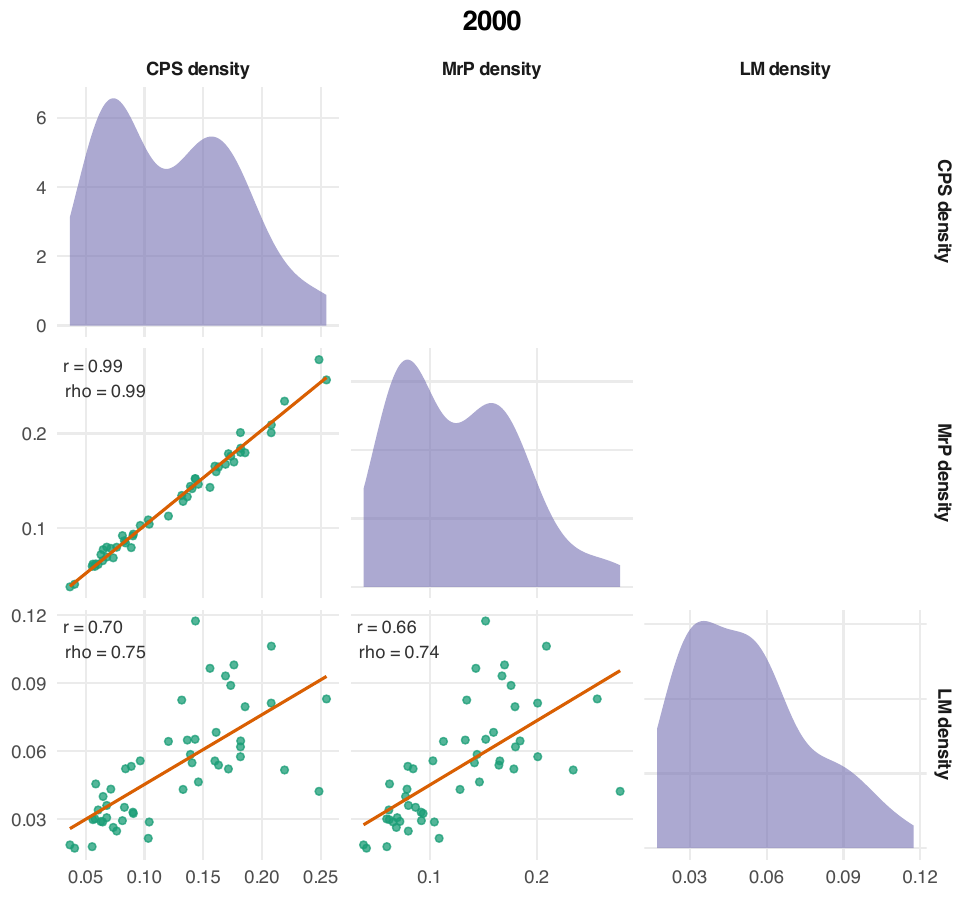}
  \caption{Union density at the state level, 2000: pairwise comparison of CPS, MrP, and LM, using the exclusive LM membership definition.}
  \label{fig:state_matrix_density_exclude_2000}
\end{figure}

\begin{figure}[htbp]
  \centering
  \includegraphics[width=\linewidth]{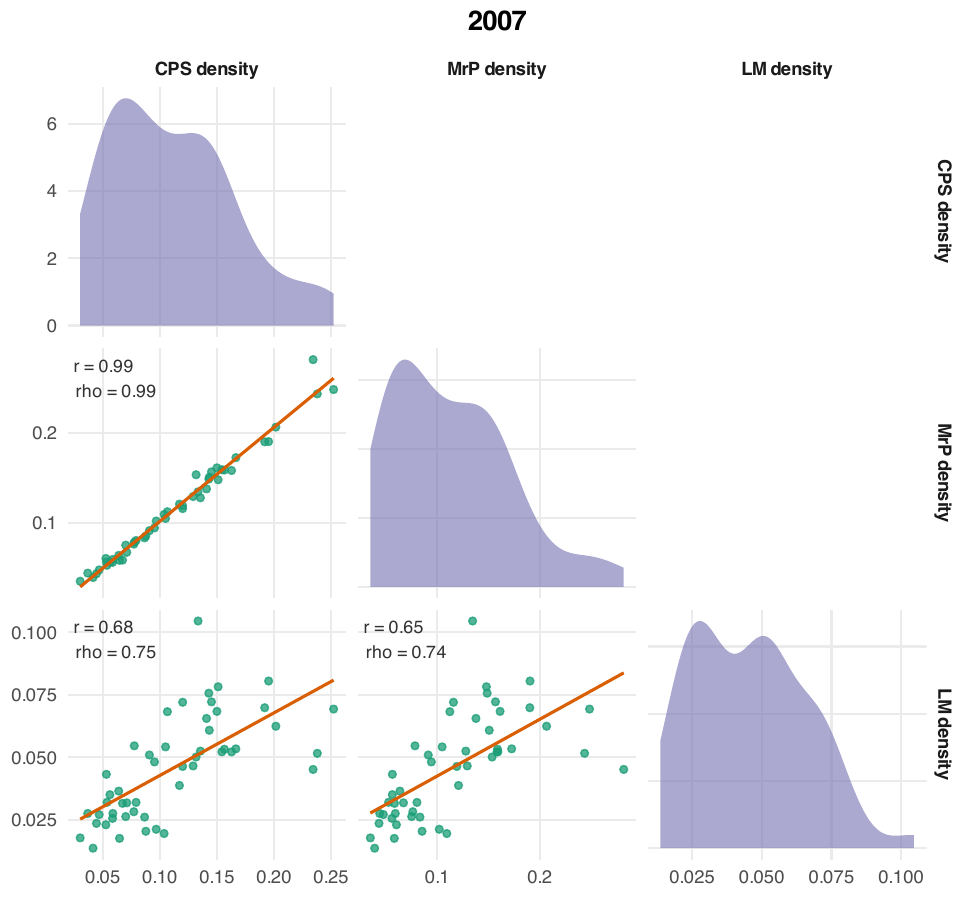}
  \caption{Union density at the state level, 2007: pairwise comparison of CPS, MrP, and LM, using the exclusive LM membership definition.}
  \label{fig:state_matrix_density_exclude_2007}
\end{figure}

\begin{figure}[htbp]
  \centering
  \includegraphics[width=\linewidth]{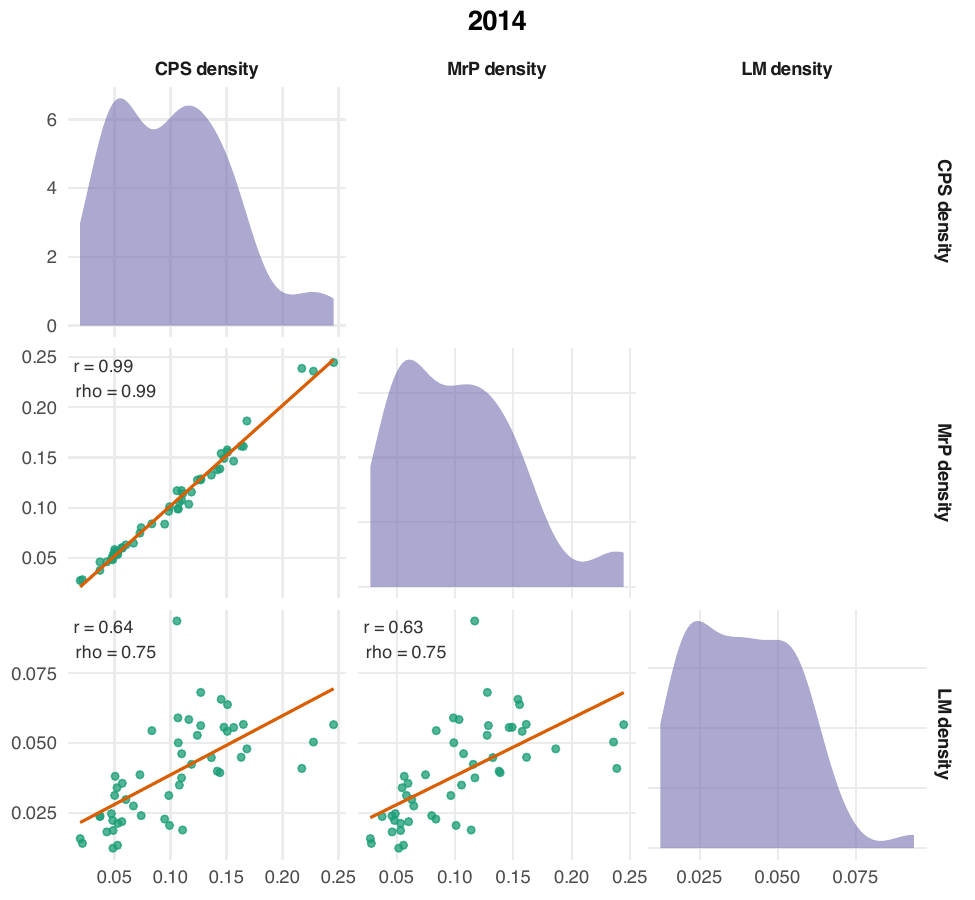}
  \caption{Union density at the state level, 2014: pairwise comparison of CPS, MrP, and LM, using the exclusive LM membership definition.}
  \label{fig:state_matrix_density_exclude_2014}
\end{figure}

\begin{figure}[htbp]
  \centering
  \includegraphics[width=\linewidth]{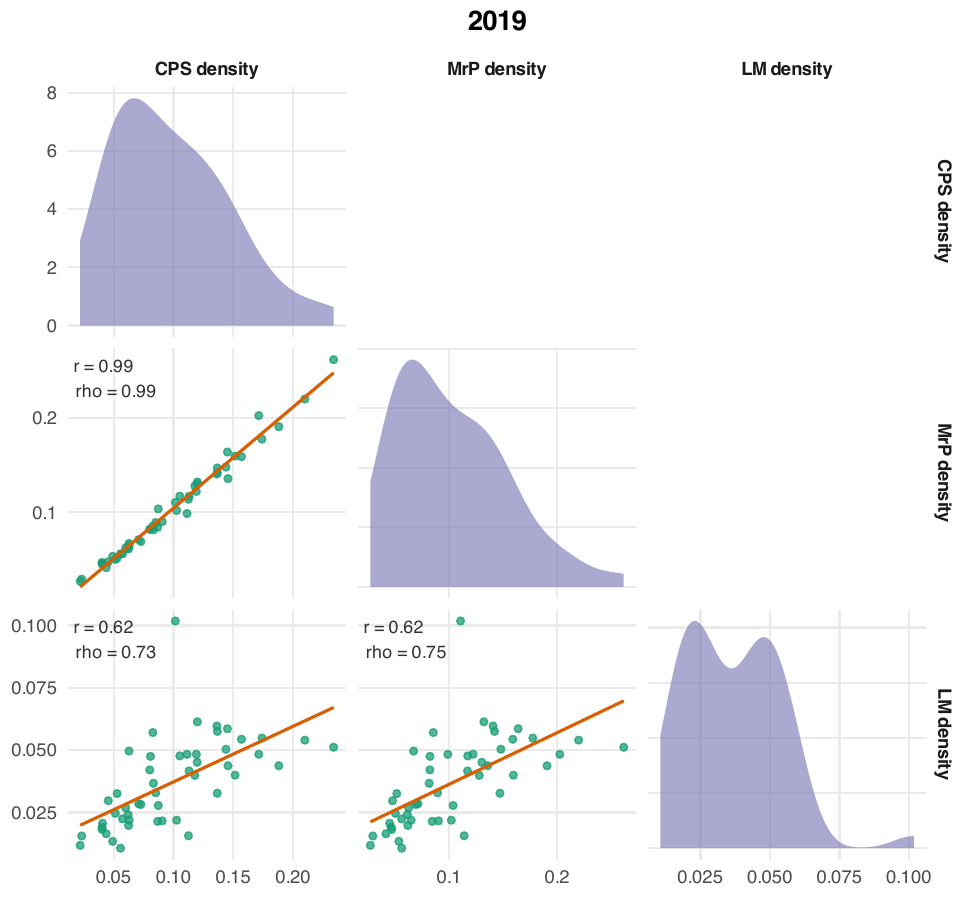}
  \caption{Union density at the state level, 2019: pairwise comparison of CPS, MrP, and LM, using the exclusive LM membership definition.}
  \label{fig:state_matrix_density_exclude_2019}
\end{figure}

\begin{figure}[htbp]
  \centering
  \includegraphics[width=\linewidth]{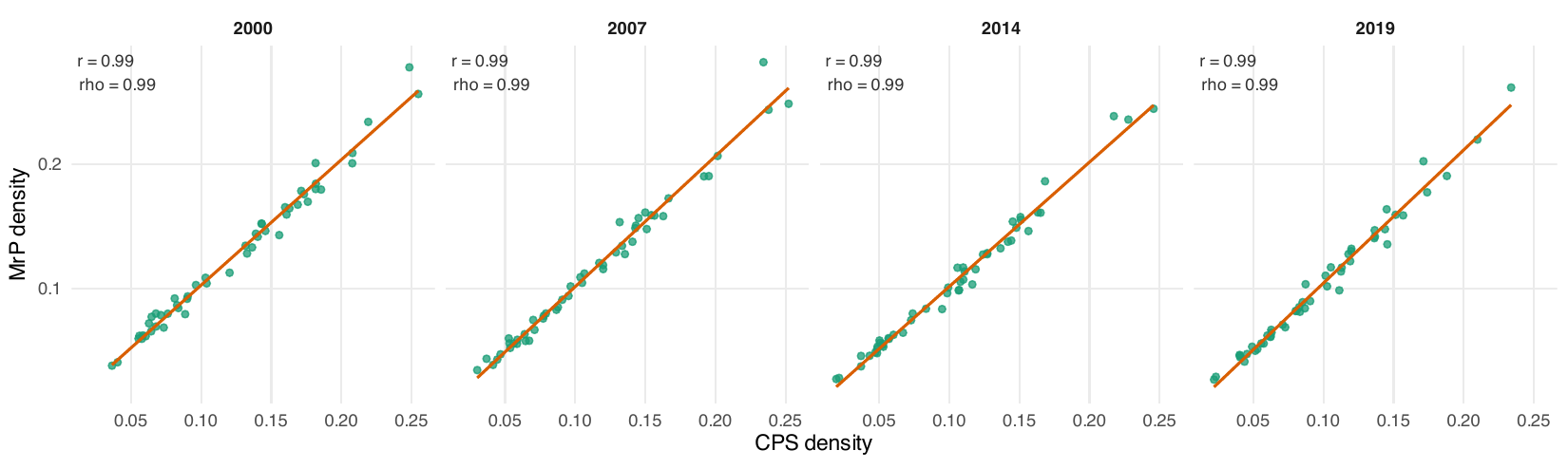}
  \caption{Union density at the state level: CPS versus MrP, faceted by year (2000, 2007, 2014, 2019), using the exclusive LM membership definition.}
  \label{fig:state_facet_density_exclude_cps_mrp}
\end{figure}

\begin{figure}[htbp]
  \centering
  \includegraphics[width=\linewidth]{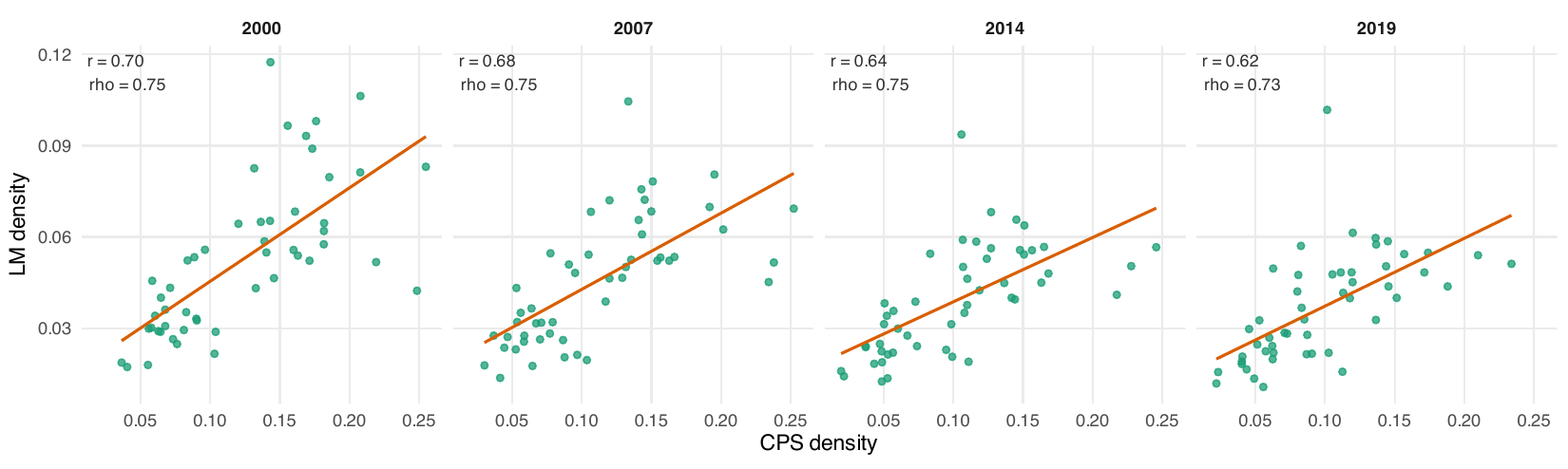}
  \caption{Union density at the state level: CPS versus LM, faceted by year (2000, 2007, 2014, 2019), using the exclusive LM membership definition.}
  \label{fig:state_facet_density_exclude_cps_lm}
\end{figure}

\begin{figure}[htbp]
  \centering
  \includegraphics[width=\linewidth]{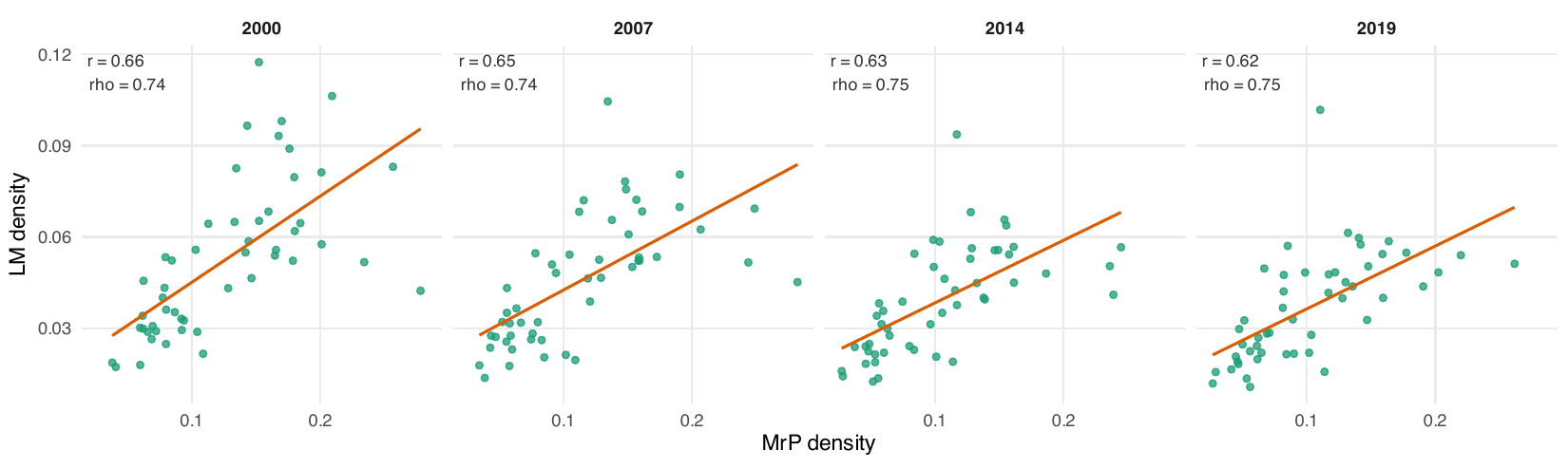}
  \caption{Union density at the state level: MrP versus LM, faceted by year (2000, 2007, 2014, 2019), using the exclusive LM membership definition.}
  \label{fig:state_facet_density_exclude_mrp_lm}
\end{figure}

\section{Within-and-between state MrP-LM (exclusive) gaps, visualized}
\label{app:district-gap-strip}

\begin{figure}[htbp]
  \centering
  \includegraphics[width=\linewidth]{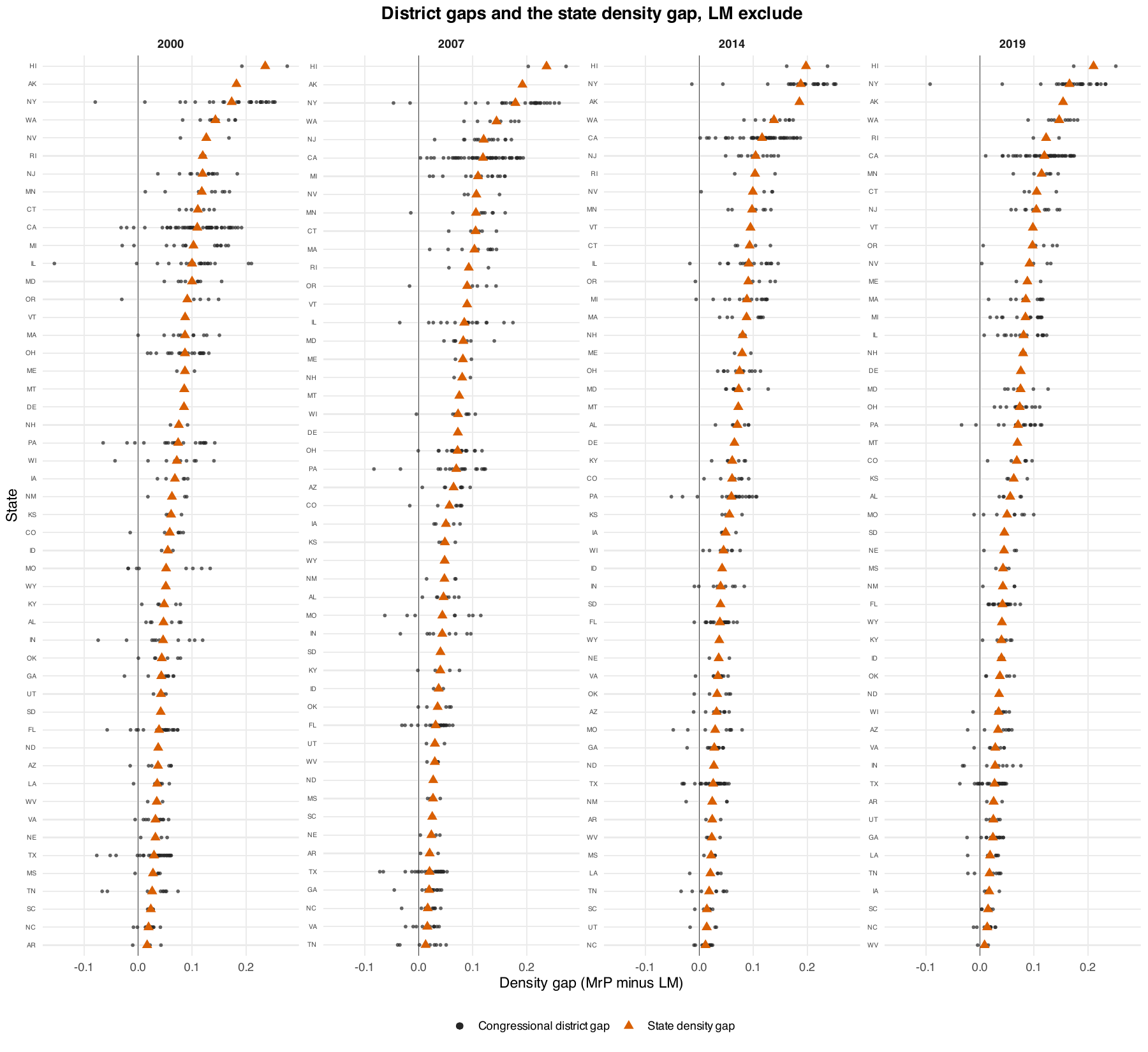}
  \caption{Congressional-district MrP--LM density gaps within each state, using the exclusive LM definition, faceted by year. Each point is a district's gap (MrP density minus LM density); the triangle marks the state density gap. States are ordered by their state density gap within each panel. DC and Louisiana~2007 excluded.}
  \label{fig:district_gap_strip}
\end{figure}
\section{Gap regressions: full specifications and diagnostics}
\label{app:gap-regressions}

The main text (Section~\ref{sec:decomp}, Table~\ref{tab:gap_reg_signed}) reports the signed gap. This appendix gives the absolute-gap analog of that table, the bare (public-sector-share-only) specification, and the county-level scatter diagnostics.

\begin{table}[htbp]
\centering
\caption{Regression of the absolute MrP--LM density gap on the state and local government employment share, with urbanicity and log employment. Absolute-gap analog of Table~\ref{tab:gap_reg_signed}. *** $p<0.001$, ** $p<0.01$, * $p<0.05$, . $p<0.1$. CB = collective bargaining.}
\label{tab:gap_reg_absolute}
\small\setlength{\tabcolsep}{5pt}
\begin{tabular}{l rr rr rr}
\toprule
 & \multicolumn{2}{c}{State} & \multicolumn{2}{c}{County} & \multicolumn{2}{c}{County $\times$ CB ban} \\
\cmidrule(lr){2-3}\cmidrule(lr){4-5}\cmidrule(lr){6-7}
 & Include & Exclude & Include & Exclude & Include & Exclude \\
\midrule
State/local gov.\ \% & -0.003 & 0.298 & 0.790* & 0.513*** & 0.898* & 0.558*** \\
 & (0.222) & (0.431) & (0.328) & (0.074) & (0.394) & (0.078) \\
$\times$ CB ban &  &  &  &  & -0.780* & -0.452*** \\
 &  &  &  &  & (0.385) & (0.090) \\
Public sector CB banned &  &  &  &  & 3.936 & -0.003 \\
 &  &  &  &  & (4.767) & (1.176) \\
Urbanicity & 0.055* & 0.162** & -0.920* & 0.030 & -1.299*** & -0.353 \\
 & (0.022) & (0.056) & (0.366) & (0.308) & (0.340) & (0.260) \\
Log employment & -1.494*** & -1.026 & 0.261 & 0.408 & -0.531 & -0.357 \\
 & (0.260) & (0.872) & (0.725) & (0.376) & (0.670) & (0.328) \\
\addlinespace
Year fixed effects & Yes & Yes & Yes & Yes & Yes & Yes \\
\addlinespace
Observations & 200 & 200 & 1,613 & 1,613 & 1,613 & 1,613 \\
Clusters (units) & 50 & 50 & 475 & 475 & 475 & 475 \\
W/in $R^2$ & 0.296 & 0.205 & 0.069 & 0.132 & 0.128 & 0.339 \\
\bottomrule
\end{tabular}

\end{table}

\begin{table}[htbp]
\centering
\caption{Bare specification: regression of the signed MrP--LM density gap on the state and local government employment share only, by geography and LM definition. *** $p<0.001$, ** $p<0.01$, * $p<0.05$, . $p<0.1$.}
\label{tab:gap_reg_bare_signed}
\begin{tabular}{l rr rr}
\toprule
 & \multicolumn{2}{c}{State} & \multicolumn{2}{c}{County} \\
\cmidrule(lr){2-3}\cmidrule(lr){4-5}
 & Include & Exclude & Include & Exclude \\
\midrule
State/local gov.\ \% & 0.236 & 0.162 & 0.313 & 0.538*** \\
 & (0.245) & (0.379) & (0.369) & (0.074) \\
\addlinespace
Year fixed effects & Yes & Yes & Yes & Yes \\
\addlinespace
Observations & 200 & 200 & 1,613 & 1,613 \\
Clusters (units) & 50 & 50 & 475 & 475 \\
W/in $R^2$ & 0.036 & 0.005 & 0.008 & 0.123 \\
\bottomrule
\end{tabular}

\end{table}

\begin{figure}[htbp]
  \centering
  \begin{subfigure}{0.48\linewidth}
    \centering
    \includegraphics[width=\linewidth]{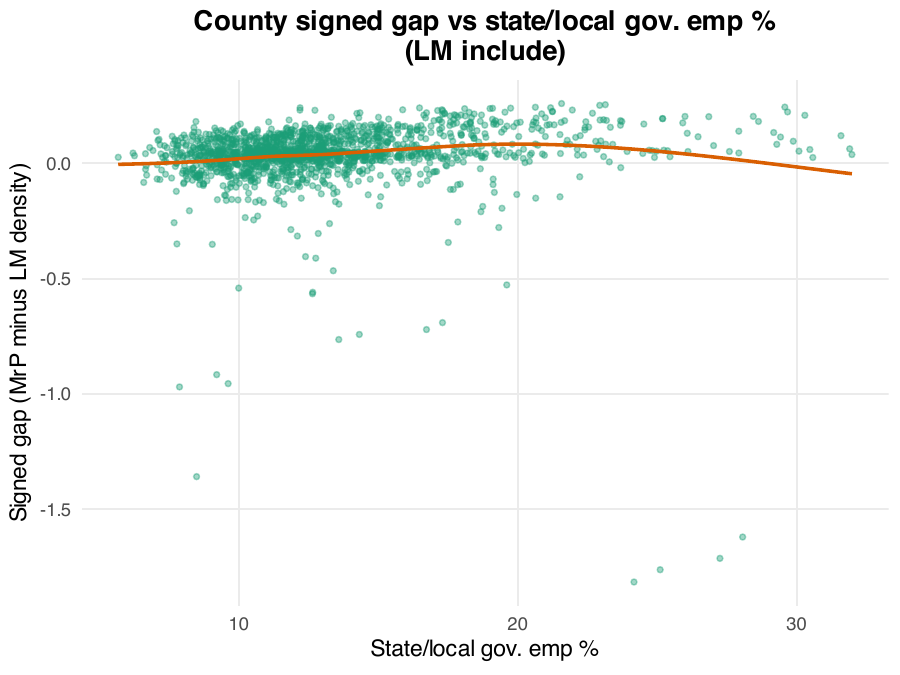}
    \caption{Signed gap, inclusive definition.}
    \label{fig:gap_scatter_signed_include}
  \end{subfigure}
  \hfill
  \begin{subfigure}{0.48\linewidth}
    \centering
    \includegraphics[width=\linewidth]{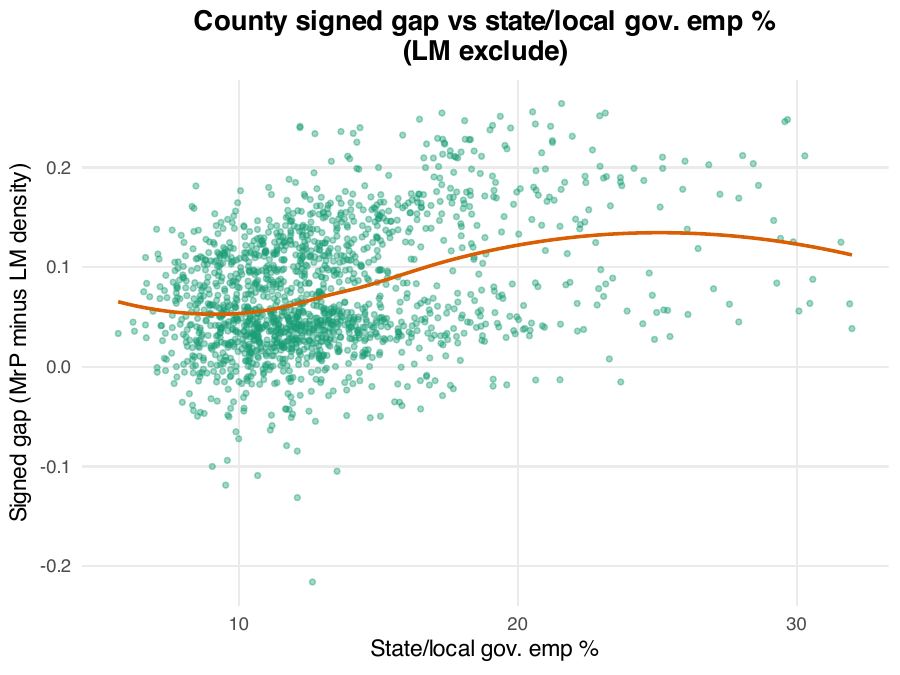}
    \caption{Signed gap, exclusive definition.}
    \label{fig:gap_scatter_signed_exclude}
  \end{subfigure}

  \vspace{1em}

  \begin{subfigure}{0.48\linewidth}
    \centering
    \includegraphics[width=\linewidth]{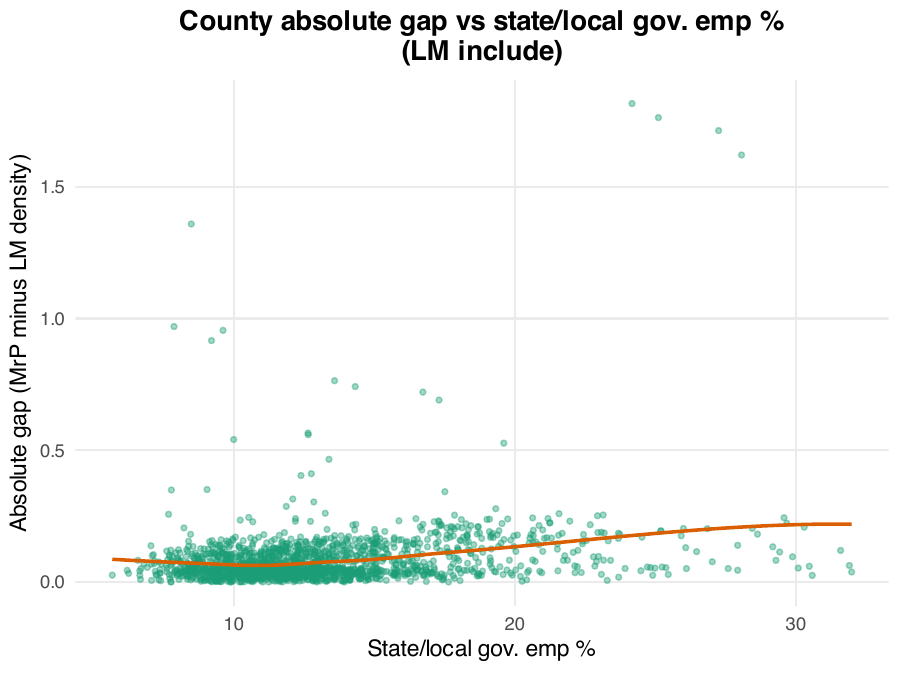}
    \caption{Absolute gap, inclusive definition.}
    \label{fig:gap_scatter_absolute_include}
  \end{subfigure}
  \hfill
  \begin{subfigure}{0.48\linewidth}
    \centering
    \includegraphics[width=\linewidth]{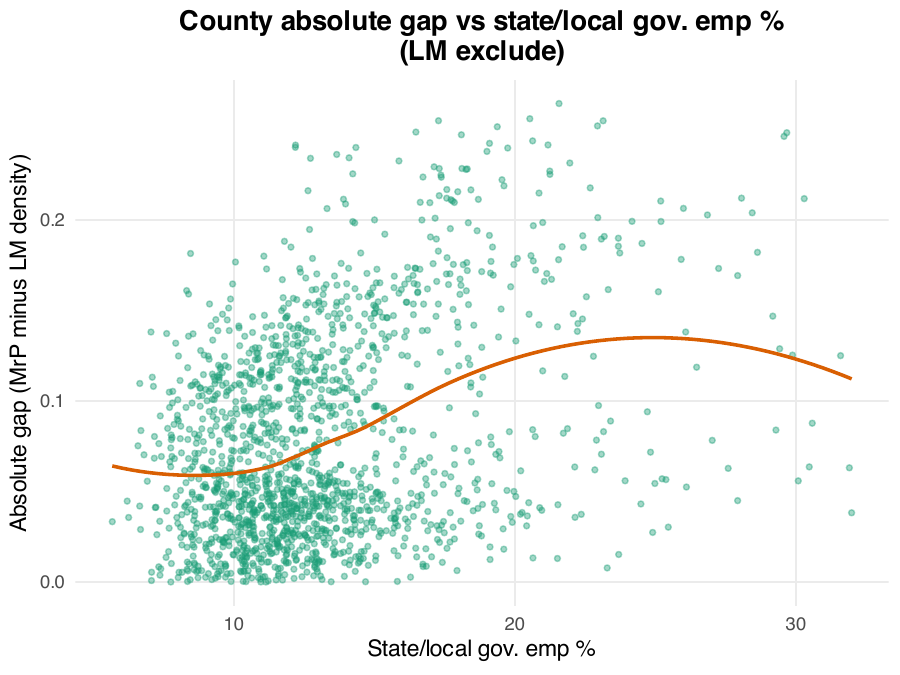}
    \caption{Absolute gap, exclusive definition.}
    \label{fig:gap_scatter_absolute_exclude}
  \end{subfigure}

  \caption{County-level MrP--LM density gap against the state and local government employment share (in percent), pooled over 2000, 2007, 2014, and 2019, with a LOESS smoother (no confidence band). Density is ACS-denominated.}
  \label{fig:gap_pubsec_scatter}
\end{figure}

\section{District-level structure of the MrP--LM gap}
\label{app:district-structure}

\begin{figure}[htbp]
  \centering
  \includegraphics[width=\linewidth]{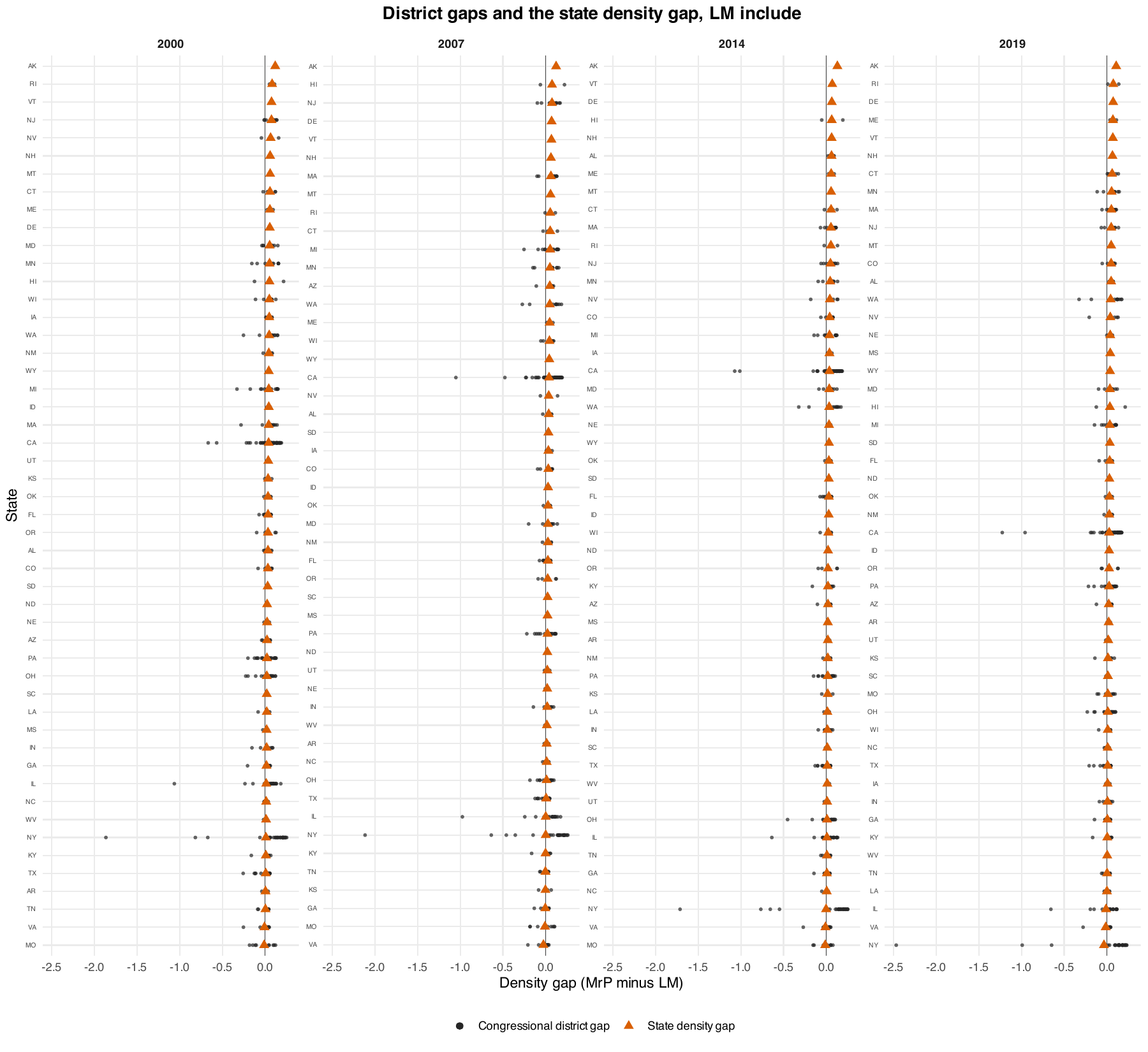}
  \caption{Congressional-district MrP--LM density gaps within each state, inclusive LM definition, faceted by year. The triangle marks the state density gap; states are ordered by it. DC and Louisiana~2007 excluded.}
  \label{fig:district_gap_strip_include}
\end{figure}

\begin{figure}[htbp]
  \centering
  \begin{subfigure}{\linewidth}
    \centering
    \includegraphics[width=\linewidth]{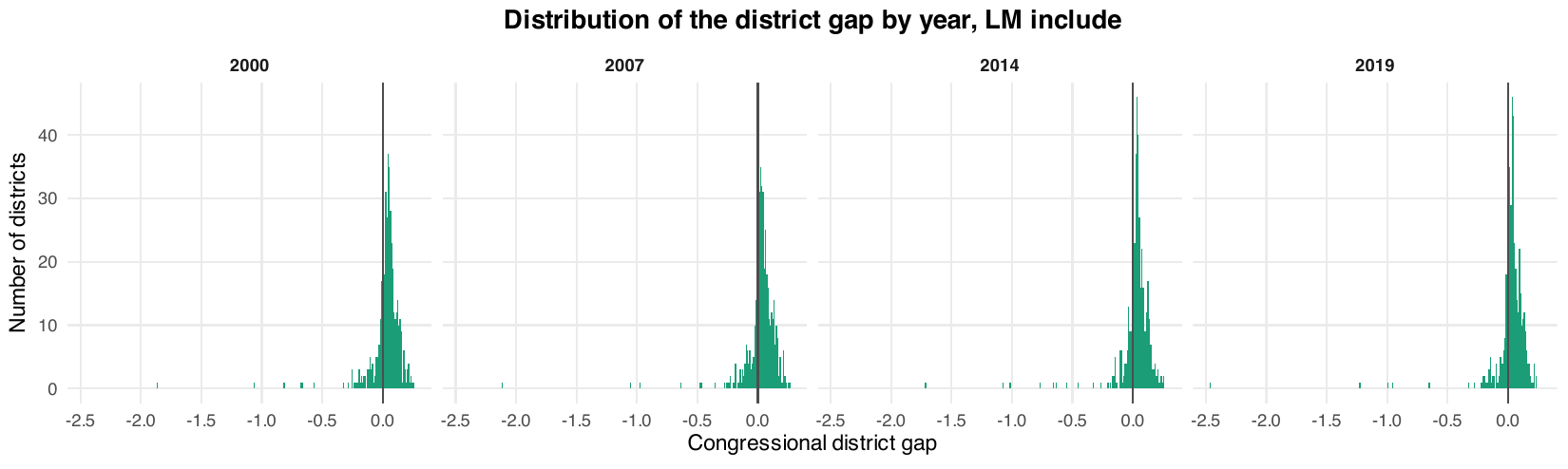}
    \caption{Inclusive LM definition.}
    \label{fig:district_gap_distribution_include}
  \end{subfigure}

  \vspace{1em}

  \begin{subfigure}{\linewidth}
    \centering
    \includegraphics[width=\linewidth]{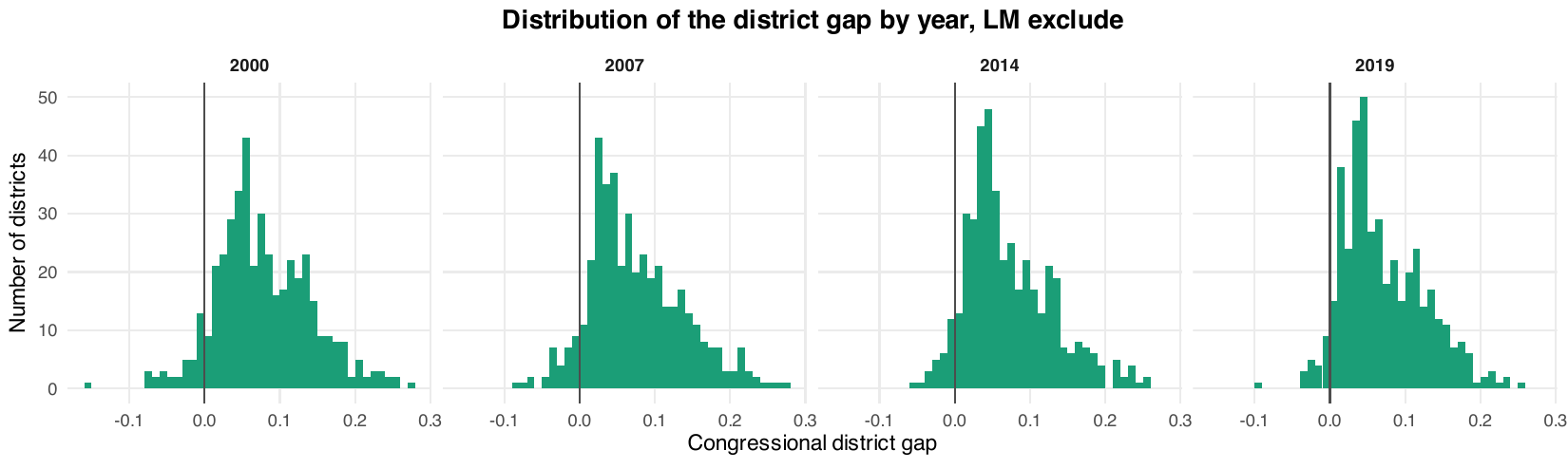}
    \caption{Exclusive LM definition.}
    \label{fig:district_gap_distribution_exclude}
  \end{subfigure}

  \caption{Distribution of the congressional-district MrP--LM density gap by year, with a reference line at zero. Density is ACS-denominated; DC and Louisiana~2007 excluded.}
  \label{fig:district_gap_distribution}
\end{figure}

This appendix supplements Section~\ref{sec:ladder-descent}. Figure~\ref{fig:district_gap_strip_include} repeats the within-state district-gap display for the inclusive definition, and Figure~\ref{fig:district_gap_distribution} shows the pooled distribution of the district gap. Tables~\ref{tab:cd_gap_reg_signed} and~\ref{tab:cd_gap_reg_absolute} carry the Section~\ref{sec:decomp} public-sector gap regression down to the district level.

\begin{table}[htbp]
\centering
\caption{Regression of the signed MrP--LM density gap at the congressional district level on the state and local government employment share, by LM definition. Urbanicity has no district equivalent and is omitted. DC and Louisiana~2007 excluded. *** $p<0.001$, ** $p<0.01$, * $p<0.05$, . $p<0.1$. CB = collective bargaining.}
\label{tab:cd_gap_reg_signed}
\begin{tabular}{l rr rr rr}
\toprule
 & \multicolumn{2}{c}{Bare} & \multicolumn{2}{c}{Controlled} & \multicolumn{2}{c}{CB ban} \\
\cmidrule(lr){2-3}\cmidrule(lr){4-5}\cmidrule(lr){6-7}
 & Include & Exclude & Include & Exclude & Include & Exclude \\
\midrule
State/local gov.\ \% & 1.114** & 0.637* & 1.145*** & 0.663* & 1.234** & 0.759** \\
 & (0.335) & (0.254) & (0.312) & (0.254) & (0.355) & (0.251) \\
$\times$ CB ban &  &  &  &  & -0.499 & -0.474 \\
 &  &  &  &  & (0.433) & (0.287) \\
Public sector CB banned &  &  &  &  & 4.209 & -0.216 \\
 &  &  &  &  & (6.309) & (3.302) \\
Log employment &  &  & 2.264 & 1.911 & 2.776 & 4.279. \\
 &  &  & (3.638) & (2.528) & (4.158) & (2.251) \\
\addlinespace
Year fixed effects & Yes & Yes & Yes & Yes & Yes & Yes \\
\addlinespace
Observations & 1,733 & 1,733 & 1,733 & 1,733 & 1,733 & 1,733 \\
Clusters (states) & 50 & 50 & 50 & 50 & 50 & 50 \\
Base w/in $R^2$ & 0.000 & 0.000 & 0.003 & 0.006 & 0.006 & 0.191 \\
W/in $R^2$ & 0.047 & 0.111 & 0.048 & 0.113 & 0.052 & 0.313 \\
$\Delta$ w/in $R^2$ & 0.047 & 0.111 & 0.044 & 0.107 & 0.046 & 0.122 \\
\bottomrule
\end{tabular}

\end{table}

\begin{table}[htbp]
\centering
\caption{Absolute-gap analog of Table~\ref{tab:cd_gap_reg_signed}: regression of the absolute MrP--LM district-level density gap on the state and local government employment share, by LM definition. *** $p<0.001$, ** $p<0.01$, * $p<0.05$, . $p<0.1$.}
\label{tab:cd_gap_reg_absolute}
\begin{tabular}{l rr rr rr}
\toprule
 & \multicolumn{2}{c}{Bare} & \multicolumn{2}{c}{Controlled} & \multicolumn{2}{c}{CB ban} \\
\cmidrule(lr){2-3}\cmidrule(lr){4-5}\cmidrule(lr){6-7}
 & Include & Exclude & Include & Exclude & Include & Exclude \\
\midrule
State/local gov.\ \% & 0.044 & 0.547* & -0.008 & 0.554* & 0.087 & 0.672** \\
 & (0.116) & (0.245) & (0.134) & (0.248) & (0.125) & (0.240) \\
$\times$ CB ban &  &  &  &  & -0.477** & -0.608* \\
 &  &  &  &  & (0.161) & (0.257) \\
Public sector CB banned &  &  &  &  & 0.067 & 1.998 \\
 &  &  &  &  & (3.017) & (2.811) \\
Log employment &  &  & -3.873 & 0.507 & -1.617 & 2.531 \\
 &  &  & (3.795) & (2.441) & (3.884) & (2.135) \\
\addlinespace
Year fixed effects & Yes & Yes & Yes & Yes & Yes & Yes \\
\addlinespace
Observations & 1,733 & 1,733 & 1,733 & 1,733 & 1,733 & 1,733 \\
Clusters (states) & 50 & 50 & 50 & 50 & 50 & 50 \\
Base w/in $R^2$ & 0.000 & 0.000 & 0.001 & 0.009 & 0.034 & 0.194 \\
W/in $R^2$ & 0.000 & 0.096 & 0.001 & 0.096 & 0.036 & 0.304 \\
$\Delta$ w/in $R^2$ & 0.000 & 0.096 & 0.000 & 0.087 & 0.002 & 0.110 \\
\bottomrule
\end{tabular}

\end{table}

\section{China-shock application: additional specifications}
\label{app:china-shock}

\begin{figure}[htbp]
\centering
\includegraphics[width=\linewidth]{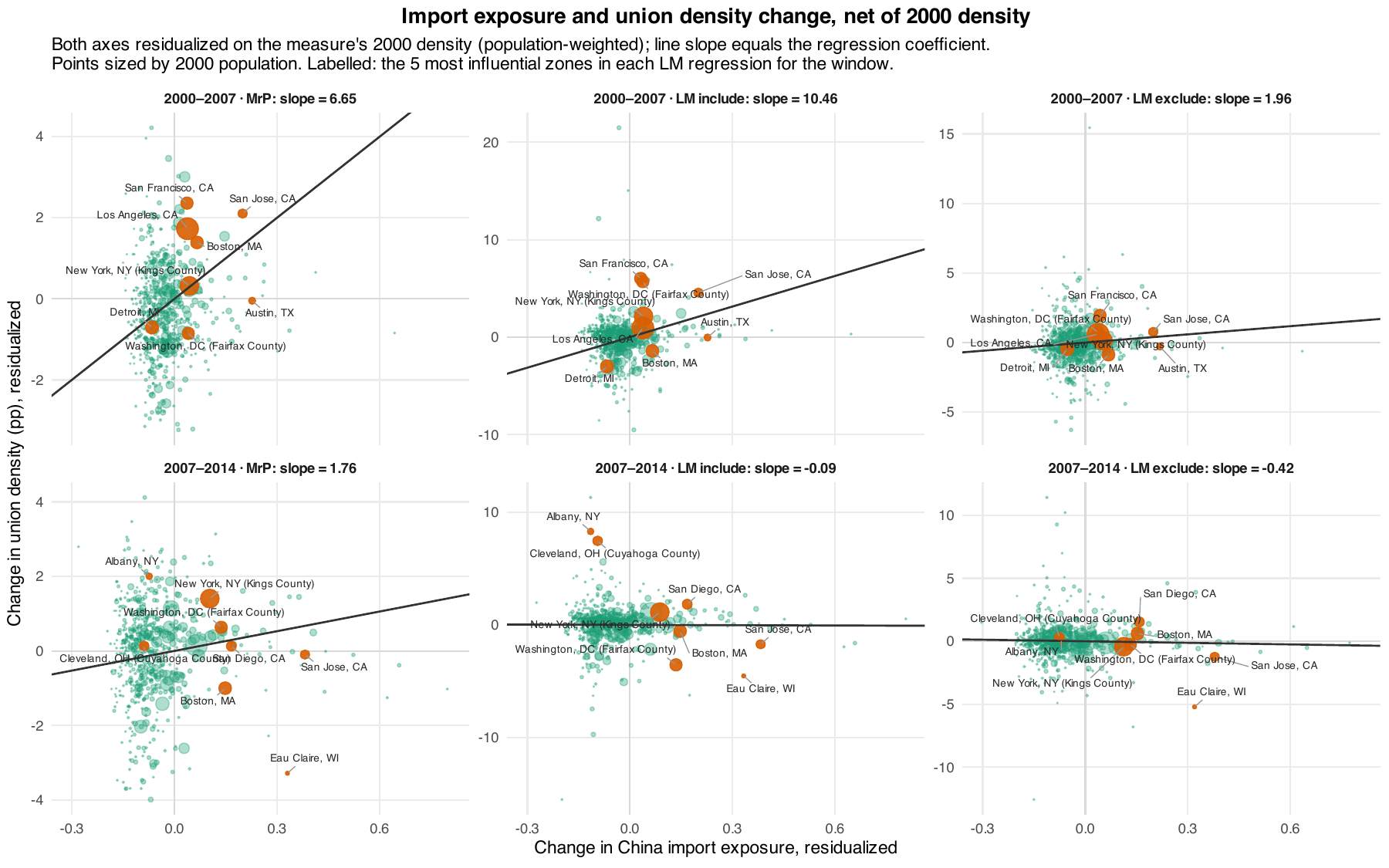}
\caption{Residualized (Frisch--Waugh--Lovell) scatter of the change in commuting-zone union density against the change in import exposure, for the main specification of Table~\ref{tab:china_shock_main}. Columns are the density measure, rows the window. Points are sized by 2000 commuting-zone population; the five most influential zones per LM regression are named.}
\label{fig:china_shock_fwl}
\end{figure}

These tables supplement Section~\ref{sec:consequences}. Table~\ref{tab:china_shock_adh_cov} repeats the main specification with covariates; Table~\ref{tab:china_shock_ps} substitutes the Pierce--Schott NTR-gap exposure measure for the ADH shift-share. Figure~\ref{fig:china_shock_fwl} shows the main specification as a residualized scatter.

\begin{table}[htbp]
\centering
\caption{China import exposure and the change in commuting-zone union density, by density measure, adding covariates (college share, skill share, capital--labor ratio, right-to-work status). Upper panel 2000--2007, lower 2007--2014.}
\label{tab:china_shock_adh_cov}
\begin{tabular}{l ccc}
\toprule
 & MrP & LM include & LM exclude \\
\midrule
\multicolumn{4}{l}{\textit{2000--2007}} \\
\addlinespace
$\Delta\mathrm{IPW}_{o}$ & 3.054* & 8.681* & 1.153 \\
 & (1.371) & (3.393) & (0.858) \\
Union density, 2000 & 0.049 & -0.020 & -0.229** \\
 & (0.031) & (0.018) & (0.032) \\
College share & -0.516 & 13.088** & 2.604* \\
 & (2.446) & (4.556) & (1.275) \\
Skill share & 9.204* & -10.622* & -1.205 \\
 & (4.292) & (4.468) & (2.016) \\
Capital-labor ratio & 0.088 & -0.203 & 0.011 \\
 & (0.245) & (0.426) & (0.141) \\
Right-to-work & 0.868* & 0.736* & 0.014 \\
 & (0.352) & (0.347) & (0.139) \\
\addlinespace
Observations & 669 & 669 & 669 \\
Clusters & 48 & 48 & 48 \\
R-squared & 0.199 & 0.180 & 0.394 \\
\addlinespace
\multicolumn{4}{l}{\textit{2007--2014}} \\
\addlinespace
$\Delta\mathrm{IPW}_{o}$ & 2.355* & 0.777 & -0.348 \\
 & (1.030) & (0.956) & (0.554) \\
Union density, 2000 & -0.044 & -0.005 & -0.142** \\
 & (0.051) & (0.048) & (0.021) \\
College share & 1.075 & -6.956 & -3.209** \\
 & (1.788) & (3.690) & (0.981) \\
Skill share & -4.918 & 4.927 & 3.945 \\
 & (4.694) & (4.351) & (1.998) \\
Capital-labor ratio & -0.024 & -0.154 & 0.260 \\
 & (0.152) & (0.348) & (0.166) \\
Right-to-work & 0.635 & 0.287 & -0.153 \\
 & (0.435) & (0.326) & (0.113) \\
\addlinespace
Observations & 669 & 669 & 669 \\
Clusters & 48 & 48 & 48 \\
R-squared & 0.234 & 0.059 & 0.226 \\
\midrule
\multicolumn{4}{@{}l@{}}{\footnotesize Standard errors clustered by state.} \\
\multicolumn{4}{@{}l@{}}{\footnotesize ** p < 0.01, * p < 0.05} \\
\multicolumn{4}{@{}l@{}}{\footnotesize 669 zones in the 48 contiguous states} \\
\bottomrule
\end{tabular}

\end{table}

\begin{table}[htbp]
\centering
\caption{China import exposure and the change in commuting-zone union density, by density measure, using the Pierce--Schott NTR-gap exposure shift-share in place of the ADH measure. Upper panel 2000--2007, lower 2007--2014.}
\label{tab:china_shock_ps}
\begin{tabular}{l ccc}
\toprule
 & MrP & LM include & LM exclude \\
\midrule
\multicolumn{4}{l}{\textit{2000--2007}} \\
\addlinespace
NTR-gap exposure & 11.529** & 15.007* & 1.991 \\
 & (2.746) & (5.834) & (1.913) \\
Union density, 2000 & -0.006 & -0.020 & -0.226** \\
 & (0.028) & (0.024) & (0.029) \\
\addlinespace
Observations & 669 & 669 & 669 \\
Clusters & 48 & 48 & 48 \\
R-squared & 0.093 & 0.050 & 0.377 \\
\addlinespace
\multicolumn{4}{l}{\textit{2007--2014}} \\
\addlinespace
NTR-gap exposure & 2.121 & 4.879* & -1.456 \\
 & (3.331) & (2.155) & (1.232) \\
Union density, 2000 & -0.085* & -0.036 & -0.134** \\
 & (0.042) & (0.042) & (0.018) \\
\addlinespace
Observations & 669 & 669 & 669 \\
Clusters & 48 & 48 & 48 \\
R-squared & 0.173 & 0.028 & 0.195 \\
\midrule
\multicolumn{4}{@{}l@{}}{\footnotesize Standard errors clustered by state.} \\
\multicolumn{4}{@{}l@{}}{\footnotesize ** p < 0.01, * p < 0.05} \\
\multicolumn{4}{@{}l@{}}{\footnotesize 669 zones in the 48 contiguous states} \\
\bottomrule
\end{tabular}

\end{table}

\end{document}